\documentclass[12pt,preprint]{aastex}
\usepackage[nolists,tablesfirst]{endfloat}

\def\deg{$^\circ$}

\def\mic{{$\mu$m}}

\def\h2o{H$_2$O}

\def\teff{$T_{\rm eff}$}

\def\mbol{$M_{\rm bol}$}

\def\aple{$\mathrel{\hbox{\rlap{\hbox{\lower4pt\hbox{$\sim$}}}\hbox{$<$}}}$}

\def\apge{$\mathrel{\hbox{\rlap{\hbox{\lower4pt\hbox{$\sim$}}}\hbox{$>$}}}$}

\begin{document}

\title{Really Cool Stars and the Star Formation History at the
Galactic Center}

\author{R. D. Blum\altaffilmark{1}} \affil{Cerro Tololo Interamerican
Observatory, Casilla 603, La Serena, Chile\\ rblum@ctio.noao.edu}

\author{Solange V. Ram\'{\i}rez\altaffilmark{1}} \affil{SIRTF Science
Center, JPL/Caltech, Pasadena, CA 91125,
USA\\solange@astro.caltech.edu}

\author{K. Sellgren\altaffilmark{1}} \affil{ Astronomy Department, The
Ohio State University, 140 West 18th Ave, Columbus, OH 43210, USA
\\sellgren@astronomy.ohio-state.edu}

\author{K. Olsen\affil{Cerro Tololo Interamerican Observatory, Casilla
603, La Serena, Chile\\ kolsen@noao.edu}}

\altaffiltext{1}{Visiting Astronomer, Cerro Tololo Interamerican
Observatory, National Optical Astronomy Observatories, which is
operated by Associated Universities for Research in Astronomy, Inc.,
under cooperative agreement with the National Science Foundation}

\begin{abstract}

We present $\lambda$/$\Delta\lambda$ $=$ 550 to 1200 near infrared $H$
and $K$ spectra for a magnitude limited sample of 79 asymptotic giant
branch and cool supergiant stars in the central $\approx$ 5 pc
(diameter) of the Galaxy. We use a set of similar spectra obtained for
solar neighborhood stars with known \teff \ and \mbol \ that is in the
same range as the Galactic center (GC) sample to derive \teff \ and
\mbol \ for the GC sample. We then construct the Hertzsprung--Russell
(HRD) diagram for the GC sample. Using an automated maximum likelihood
routine, we derive a coarse star formation history of the GC.  We find
(1) roughly 75$\%$ of the stars formed in the central few pc are older
than 5 Gyr; (2) the star formation rate (SFR) is variable over time,
with a roughly 4 times higher star formation rate in the last 100 Myr
compared to the average SFR; (3) our model can only match dynamical
limits on the total mass of stars formed by limiting the IMF to masses
above 0.7 M$_\odot$. This could be a signature of mass segregation or
of the bias toward massive star formation from the unique star
formation conditions in the GC; (4) blue supergiants account for 12
$\%$ of the total sample observed, and the ratio of red to blue
supergiants is roughly 1.5; (5) models with isochrones with [Fe/H] =
0.0 over all ages fit the stars in our HRD better than models with
lower [Fe/H] in the oldest age bins, consistent with the finding of
Ramirez et al. (2000) that stars with ages between 10 Myr and 1 Gyr
have solar [Fe/H].

\end{abstract}
\keywords{Galaxy: center --- stars: late-type --- stars: AGB and post-AGB --- supergiants}


\section{INTRODUCTION}

The properties of the stellar population at the Galactic Center (GC)
suggest that the nucleus is distinct from the other main structural
components of the Galaxy (the Galactic disk, bulge, and Halo), though
each of these components may contribute to the integrated
population. We would like to distinguish between extensions of these
populations and a unique GC population which has formed and evolved
there.  OH/IR stars distributed in an inner disk between $\sim$1 and
100 pc \citep*{lhw92} show higher rotational velocities than expected
for a ``hot'' bulge component, suggesting a disk like population. Near
infrared surface brightness measurements indicate the bright nucleus
joins the bulge discontinuously at a radius of about 150 pc; see the
discussion given by \citet{kent92} which relies in part on the 4 \mic
\ minor axis surface brightness profile presented by \citet{lp85}.
Recent work on detailed abundance determinations in the central 60 pc
\citep{csb00,ram00} also reveal differences between the Galaxy's
nucleus and bulge components. \citet{ram00} find a narrow distribution
in [Fe/H] in the GC peaked around the solar value, while the bulge has
a very broad distribution in [Fe/H] with a mean less than the solar
value \citep{mr94,sad96}.

Stars as young as \aple 5 Myr are now known to exist in the central
pc. Very recent star formation was clearly established by
\citet{for87} and \citet*{ahh90} with the discovery of a bright,
evolved, and massive emission--line star (the ``AF''
star). \citet{keal91} further showed that a significant component of
the strong recombination lines of H and He seen toward the GC arises
in spatially compact sources, particularly the ``IRS 16'' cluster of
massive stars
\citep{naj94,lpfa95,blum95,krabbe95,blum95b,tam96,naj97}. \citet{krabbe95}
have modeled the IRS~16 cluster as the evolved descendants of the most
massive stars($\sim$ 100 M$_{\sun}$) belonging to a \aple 7 Myr old
burst.

In their review of the global phenomena on--going in the GC region,
\citet{ms96} described the properties of the ``central molecular
zone,'' or CMZ. The CMZ is a ``disk'' of enhanced molecular density
about 200 pc in radius centered on the GC. The gas is confined to a
region near the plane of the Galaxy, but with significant
non--circular motions. The distribution and presence of molecular gas
in the CMZ may in large part be due to the effects of the inner
Galactic stellar bar
\citep{liszt80,ml86,bin91,bs91,w94,dwek95,stan94}. The material in the
CMZ is fueling current star formation on this large scale at a rate of
about 0.5 M$_{\sun}$~yr$^{-1}$ \citep{gus89}, but it may also be the
ultimate source of material which is processed into stars within a few
pc of the GC \citep{ms96}. If so, angular momentum losses must funnel
the gas down to the circumnuclear disk (CND) at radii between $\sim$ 2
and 8 pc (see the extensive reviews by Genzel, Hollenbach, \& Townes
1994 and Morris \& Serabyn 1996). This molecular structure is probably
not a long--lived one, but rather periodically forms and supplies the
GC with star forming material through instabilities which cause
material to fall from its inner radius into the central pc
\citep{san99}; at present the CND may be accreting about
0.5$\times$10$^{-2}$ M$_{\sun}$~yr$^{-1}$ \citep{gus87,jack93}.

In this paper, we continue the exploration of the stellar content of
the central few pc of the Milky Way begun by \citet*[hereafter
BDS96]{bds96} and \citet*[BSD96]{bsd96}.  Using $J$, $H$, and $K$
photometry, BDS96 identified a bright component to the dereddened
$K-$band luminosity function relative to the Galactic bulge population
seen toward Baade's Window (BW), which is predominantly old (\apge~10
Gyr; Terndrup 1988; Lee 1992; Holtzman et al. 1993). Specifically,
BDS96 compared $K-$band counts in the central few pc to those in BW
presented by \citet{tiede95}. BSD96 presented a small sample of near
infrared spectra which they used to begin a detailed investigation of
the properties of the cool stellar population in the Galactic center
(GC), including the ages of individual stars which trace multiple
epochs of star formation there. This work, in turn, is built on
earlier work, most notably that of \citet*{lrt82} who investigated
recent star formation in the GC using the luminous and young M--type
supergiants they identified there.

Our goal is to determine \teff \ and \mbol \ for a magnitude limited
sample of GC stars, using the two-dimensional classification provided
by the measured strengths of the CO and \h2o \ absorption features
present in modest resolution $K-$band and $H-$band spectra. The
technique, described in \S3, is calibrated using a sample of
comparison stars with known \teff \ and \mbol, selected from the
literature to match the \teff \ and \mbol \ of the GC stars.  After
\teff \ and \mbol \ are determined for the GC stars, we place the GC
stars in the HR diagram (\S4.1) and use this to constrain the star
formation history (SFH) within the central few pc of our Galaxy. The
SFH calculation is described in \S4.2 and discussed in \S5. A brief
summary is given in \S6.

\section{OBSERVATIONS AND DATA REDUCTION}

Spectroscopic observations of the comparison and GC stars were made
using the facility IRS and OSIRIS\footnote{OSIRIS is a collaborative
project between the Ohio State University and CTIO. OSIRIS was
developed through NSF grants AST 9016112 and AST 9218449.}
spectrometers mounted on the Cerro Tololo Interamerican Observatory
(CTIO) 4m Blanco telescope over several runs beginning in 1997 and
ending in 2000 (see Tables~1 and 2). In addition, comparison stars
were observed at the Michigan--Dartmouth--MIT (MDM) 2.4m telescope on
Kitt Peak using the Ohio State University MOSAIC infrared
camera/spectrometer (Table~1).  The IRS, OSIRIS, and MOSAIC are
described by \citet{deal90}, \citet{dabfo93}, and \citet{peat98},
respectively. The IRS employed a 0.7$''\times12.5''$ slit, OSIRIS a
1.2$''\times30''$ slit, and MOSAIC a 0.6$''\times150''$ slit. The
detector pixel scales are 0.32$''$ pix$^{-1}$, 0.40$''$ pix$^{-1}$,
and 0.30$''$ pix$^{-1}$ for the IRS, OSIRIS, and MOSAIC,
respectively. The IRS and OSIRIS were used in cross--dispersed mode
giving essentially full coverage of the $J$, $H$, and $K$
bands. MOSAIC was used in $JHK$ grism mode (1.22 \mic \ to 2.29 \mic,
where the 2.29 \mic \ cutoff is due to the $JHK$ blocking/order
sorting filter). For the MOSAIC spectra, an extra, independent
$K-$band segment was obtained for each star covering the red portion
of the $K-$band. This was accomplished using the same setup, but with
a $K$ filter instead of the JHK blocker.  The spectral resolutions are
approximately $\lambda/\Delta\lambda = 1200$, 560, and 750 for OSIRIS,
the IRS, and MOSAIC, respectively.

Observing conditions varied over the course of different observing
runs. Data were obtained in photometric and non--photometric
conditions. The $K-$band seeing at the CTIO 4m was typically between
0.5$''$ and 1$''$. At the MDM 2.4m no effort was made to keep the
bright comparison stars in good focus, and in fact, sometimes the
telescope was intentionally defocused (see below).
 
All basic data reduction was accomplished using IRAF\footnote{ IRAF is
distributed by the National Optical Astronomy Observatories, which are
operated by the Association of Universities for Research in Astronomy,
Inc., under cooperative agreement with the National Science
Foundation.}. Each spectrum was flat--fielded using dome flat field
images and then sky subtracted using a median combined image formed
from the data themselves or from a set of independent sky frames
obtained off the source (for the GC stars, typically on dark clouds
$\sim$ 30$''$ to 90$''$ away). Nearby sky apertures ($\sim$ 1--2$''$
on either side of the object) were defined on the long slit images and
used to correct for over/under subtraction of the night--sky OH lines
and the unresolved background light in the case of the GC stars. For
the comparison stars, the situation varied depending on the brightness
of the star. Some of the stars had to be defocused and/or placed on
the edge of the slit in order not to saturate the detector. In a
number of cases at the 4m Blanco telescope, the mirror covers were
partially closed.  These procedures typically produced considerable
wings to the PSF. For the IRS, the slit was not long enough in such
cases to provide blank sky, so sky frames were obtained off the
source, some 5$''$ to 10$''$ away. These special procedures do not
affect the spectral resolution as confirmed by comparing the night sky 
line widths with similar spectra taken under normal conditions.

Following sky subtraction, the object spectra were extracted from the
long slit images by summing the dispersed light over $\pm$ 3--5
spatial pixels (depending on seeing and source crowding in the GC) and
then divided by the spectrum of an O, B, or A star to correct for the
telluric absorption. For the case of the bright comparison stars which
were defocused, the extraction apertures were up to approximately
$\pm$ 10 pixels. Brackett absorption features in the telluric
standards were ``fixed'' by drawing a line across the feature from
continuum points on either side.

The wavelength dispersion solutions were determined from OH lines
observed in the $H$ and $K$ bands and the line positions given by
\citet{oo92}. The wavelength zero point was set by moving the observed
position of the CO 2.3 \mic \ bandhead to 2.2935 \mic \ \citep{kh86}.

These spectra were then multiplied by a $\lambda^{-4}$ spectrum
approximating the blackbody curve for the telluric standards. For the
GC stars, a correction was made for the interstellar reddening
assuming the extinction law given by \citet{m90} and the derived
$A_K$ from BDS96.

\section{SPECTRAL CLASSIFICATION}

\subsection{Description of the Technique}

BSD96 used $K-$band spectroscopic indices for CO and \h2o \ to provide
a two--dimensional classification yielding \teff \ and \mbol \ for
cool, luminous GC stars (following Kleinmann \& Hall 1986; see also
Ram\'{\i}rez et al. 1997). The CO index is a measure of the strength
of the 2.2935 \mic \ 2--0 $^{12}$CO rotational--vibrational bandhead
\citep{kh86}. The \h2o \ feature is a broad depression of the
continuum between the $H$ and $K-$bands due to myriad blended steam
absorption lines. There are similar steam absorption bands between the
$J$ and $H$ bands and between the $K$ and $L$ bands \citep{s78}. The
latter band can affect the CO bandhead region for stars with extreme
\h2o \ absorption strength (see Sec. 3.2.1).  The combination of \h2o
\ and CO features has been used in the past to break the degeneracy in
luminosity class and \teff \ vs absorption strength which exists for
each feature alone (see the extensive discussion in \citet{kh86} and
BSD96). The correlation of band strength with luminosity is positive
for CO (CO increases in stars of higher luminosity) and negative for
\h2o \ (\h2o \ decreases for stars of higher luminosity).

Several improvements have been made in the present work, relative to
the analysis in BSD96. Both $H$ and $K$ band coverage are used to
define the \h2o \ absorption where as BSD96 had only $K-$band
spectra. In that case, the derived \h2o \ strength was sensitive to
the interstellar extinction and reddening for any given star since the
relative depression of the blue end of the $K$ band depends
sensitively on $A_K$ and the assumed wavelength ($\lambda^{-1.7}$)
dependence of the reddening law. Using both $H$ and $K-$band spectra,
the \h2o \ absorption is seen as a broad ``feature'' spanning the blue
end of the $K$ band and the red end of the $H$ band (as well as a
downturn in stellar flux at the blue end of the $H$ band and the red
end of the $K$ band).  By combining spectra of both the $H$ and $K$
bands, in continuum regions which are not affected by stellar steam
absorption, the intrinsic stellar steam absorption at 2.07 $\mu$m can
be distinguished from the interstellar reddening which produces a
monotonic decrease in flux toward bluer wavelengths.  In the present
paper, the CO vs \teff \ relation has also been improved, placing it
on a more quantitative basis which relies less on spectral types and
more on \teff \ measurements.  In terms of sample size, we are
presenting a GC sample more than three times larger than the one
presented by BSD96. A larger sample is crucial for constraining the
SFH through theoretical models.

\begin{figure}
\plotone{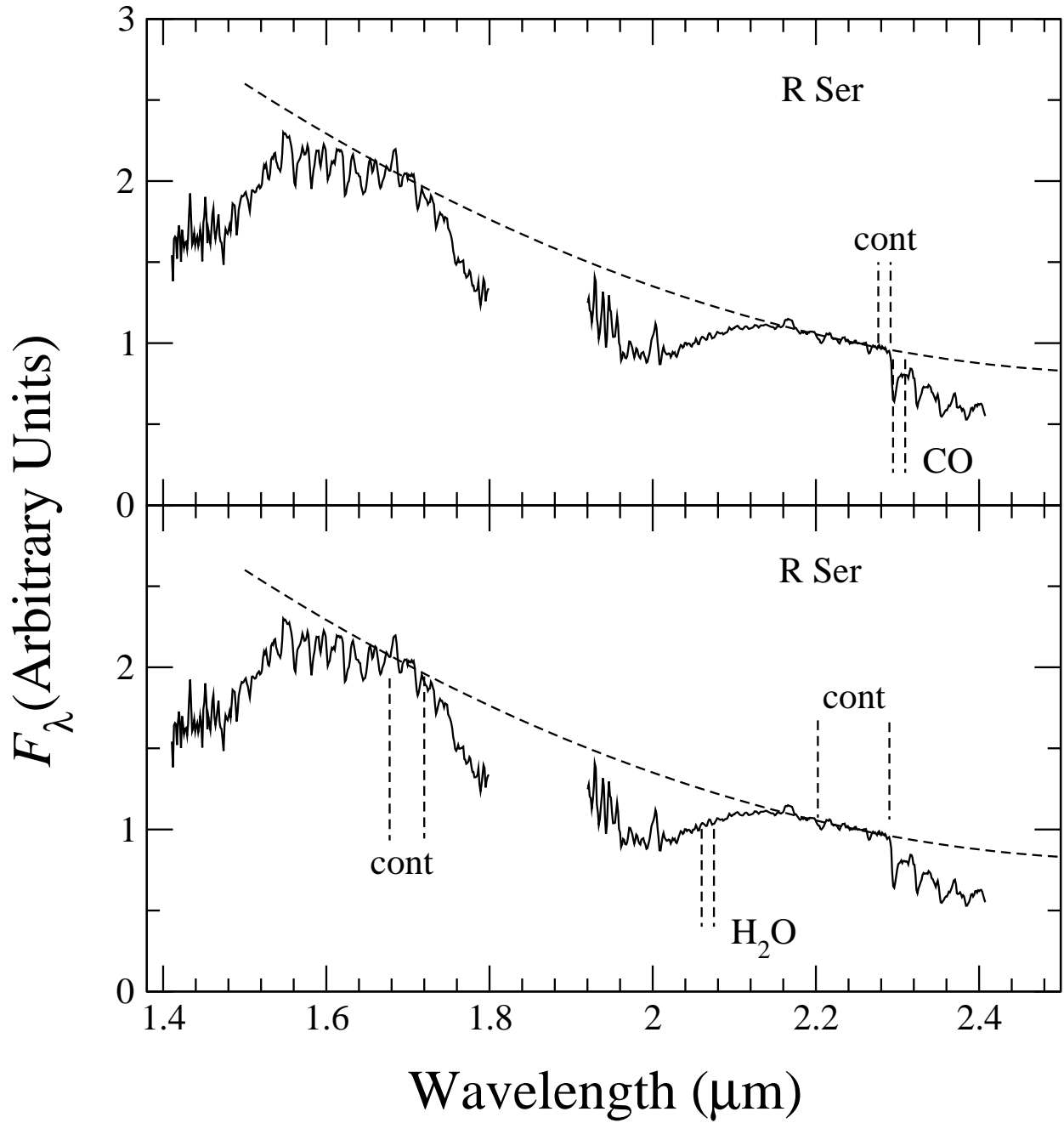} 

\caption{$H$ and $K$ spectra of the asymptotic giant branch (AGB)
star R~Ser (M7~III, Mira) used to demonstrate the CO and \h2o \
measurements for all spectra.  The CO strength is determined by the ratio
of flux in the band centered at 2.302 \mic \ compared to the flux in
the continuum band (in the star, not the fit) at 2.2875 \mic \ (the
bands are indicated with vertical {\it dashed} lines in the {\it
upper} panel).  The {\it dashed} curves are quadratic fits to the
continuum in bands at 1.68--1.72 \mic \ and 2.20--2.29 \mic \ (as
indicated in the {\it lower} panel). The \h2o \ strength is measured
using the flux in a band at 2.060--2.075 \mic \ relative to the flux
in the fit at the same position (as indicated in the {\it lower}
panel). 
\label{lpv}}
\end{figure}

The CO index is defined (BSD96) as the percentage of flux in the CO
2.3 \mic \ feature relative to a continuum band centered at 2.284 \mic
\ ([1 $-$ $F_{band}/F_{cont}] \times 100$). The CO band and continuum
band were 0.015 \mic \ wide and the CO band was centered at 2.302
\mic. The CO index is only marginally affected by extinction since the
CO and continuum bands are closely spaced. A typical CO strength of
20$\%$ changes by about 1$\%$ for a change in $A_K$ of one magnitude. This
is similar to the typical uncertainty in the derived CO strength which
is taken as the one--$\sigma$ uncertainty in feature strength derived
from the pixel--to--pixel variation in the nearby continuum.  The CO
and associated continuum band are graphically represented in the {\it
upper} panel of Figure~\ref{lpv}\footnote{The data at wavelengths
between 1.80 and 1.92 \mic \ have been omitted from the plots in
Figures~1 and 3,4,7--10 due to the low telluric transmission in these
regions.}.

The \h2o \ strength is defined similarly to the CO index, but with a
quadratic fit to the continuum using bands at 1.68--1.72 \mic \ and
2.20--2.29 \mic \ (see the {\it lower} panel of Figure~\ref{lpv}) and
a band 0.015 \mic \ wide centered at 2.0675 \mic \ (indicated in the
{\it lower} panel of Figure~\ref{lpv}).  The difference between this
index and that used by BSD96 is that the latter index used the same
continuum band as the CO index (hence the sensitivity in that work to
the adopted extinction for any given star). The formal
uncertainty in the \h2o \ strength measured here is a fraction
of a percent; the actual uncertainty, 3$\%$, is derived from the
scatter of \h2o \ measurements for supergiants, which have
no measurable \h2o.  Small changes in the choice of continuum
bands used for the fits can lead to systematic changes in the
derived \h2o \ strength of \aple 5-10$\%$, but these tend to affect 
all the spectra similarly. The systematic uncertainty should
be much smaller than that given by BSD96 since the continuum
fit spans the $H$ and $K$ bands and is thus insensitive to the
details of the reddening.

\subsection{Comparison Stars}

The comparison star list is given in Table~1. These stars were chosen
to span the range of expected GC star \mbol \ and \teff \
\citep[BSD96;][]{csb00,ram00}. The literature was surveyed for cool
giant, AGB, and cool supergiant stars with derived \teff \ and
\mbol. References for these two quantities are given in Table~1. In
addition, we have used five digital spectra previously presented by
BSD96 and \citet{lv92} for several comparison stars which we were not
able to re--observe (see Table~1) but which met our criteria on \mbol
\ and \teff. We observed two stars (FL~Ser, Z~UMa) which we later
removed from our sample because they had discrepant \teff \ compared
to stars of similar spectral type \citep*{dy98}.

\subsubsection{CO Index}

The comparison stars were used to define a CO index vs. \teff \
relation which we then used to determine the \teff \ of the GC stars
(see below). In order to produce a relationship between CO and \teff,
comparison stars were selected with fundamental, e.g. \citet{dy96},
\teff \ determinations, or \teff \ determined from detailed
spectroscopic analysis. Unfortunately, the list of such stars matching
the GC \teff \ range and also with \mbol \ determined was too small
(only four supergiants). Thus, for a number of stars in Table~1 the
\teff \ vs spectral type relation of \citet{dy98} was used to extend
our sample of comparison stars (mainly for the supergiants). A small
offset in \teff \ ($-$400~K for K type stars, and $-$220~K for M type)
was made for supergiants \citep{dy96}. \teff \ as determined from the
relationship in \citet{dy98} was substituted for two stars whose
independent \teff \ published in the literature gave substantially
larger residuals relative to the derived CO vs \teff \ relationship
(3~Cet, \citealt{lb80}, and CD--60~3621, \citealt{lb89}).

Following \citet{dy96}, two stars of luminosity class II were treated
as giants (XY Lyr and $\pi$ Her), and this is consistent with their
measured CO values compared to similar stars of the same spectral
type.  Conversely, HD163428 (luminosity class II) was included in the
supergiant category as its CO is consistent with that group. We made
analogous assignments for the S--stars shown in Table~1 which have no
luminosity class given explicitly. In all cases, the associated
$M_{\rm bol}$ is consistent with the luminosity class chosen, though
there is overlap between luminous giants (AGB stars) and less luminous
supergiants (see the extensive discussion in BSD96).

\begin{figure}
\plotone{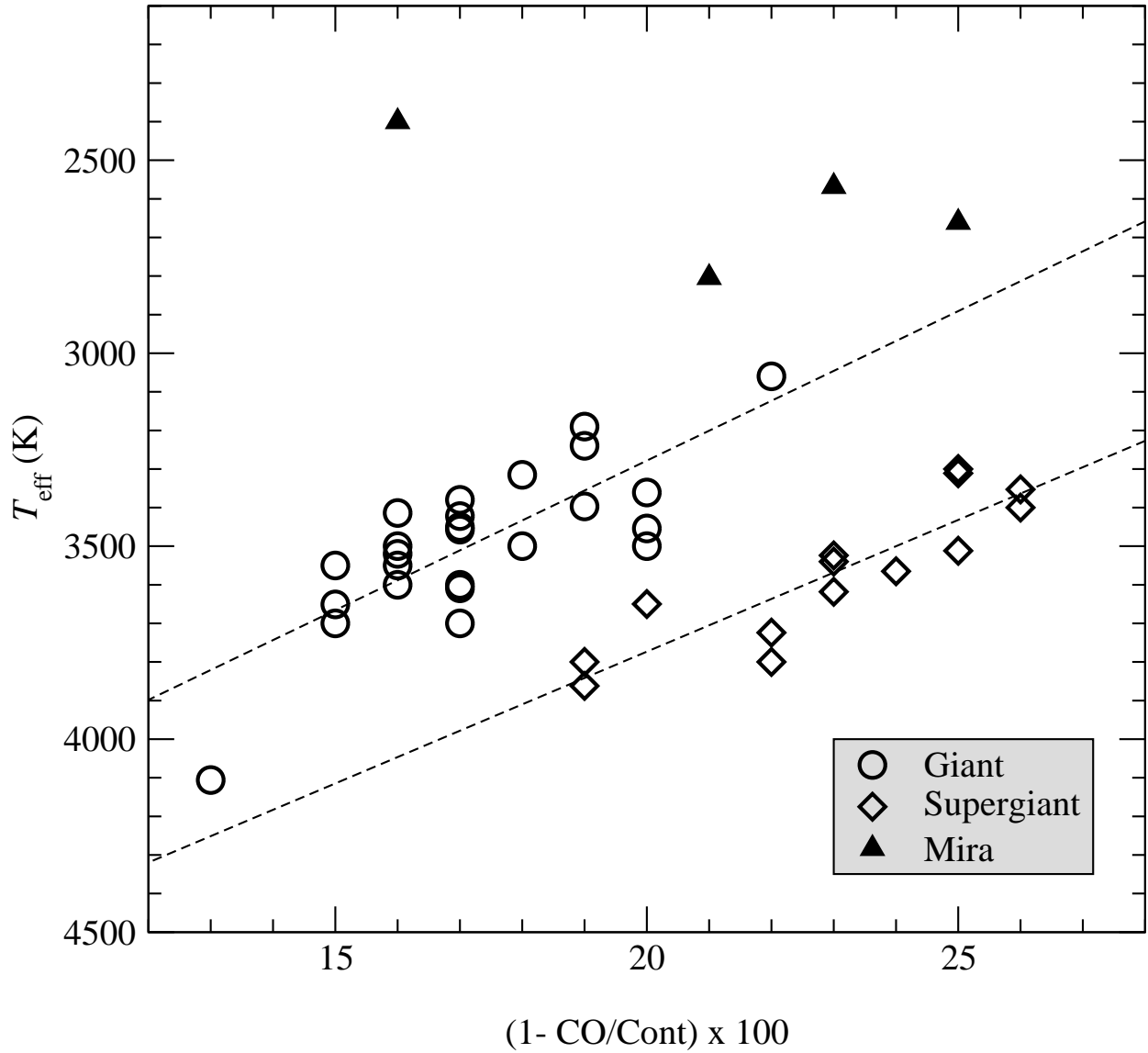} 

\caption{2.3 \mic \ CO absorption strength for the comparison
stars. CO strength increases with decreasing \teff \ but also with
increasing luminosity; see text. The correlation appears to break down
for some long period variables (Miras); see text.
\label{covteff}}
\end{figure}

\begin{figure}
\plotone{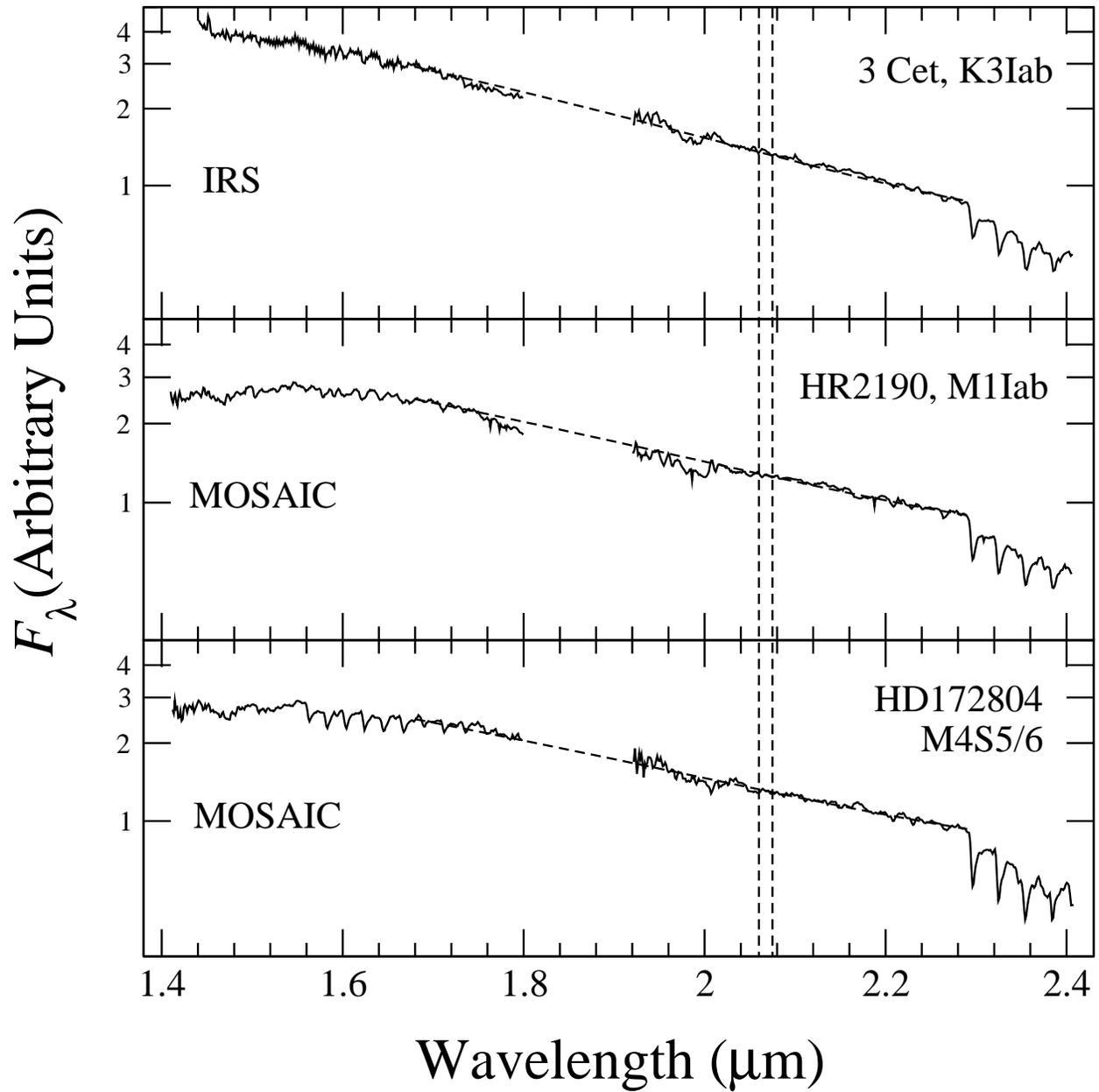} 

\caption{$H$ and $K$ spectra of late--type supergiants. Note the
increase in CO (2.3 \mic) absorption strength for later types. The
{\it dashed} curves are the fits to the continua used to measure \h2o
\ at the position of the vertical {\it dashed} lines (see text and
Figure~\ref{lpv}). Y axis scaled is logarithmically.
\label{sg}}
\end{figure}

\begin{figure}
\plotone{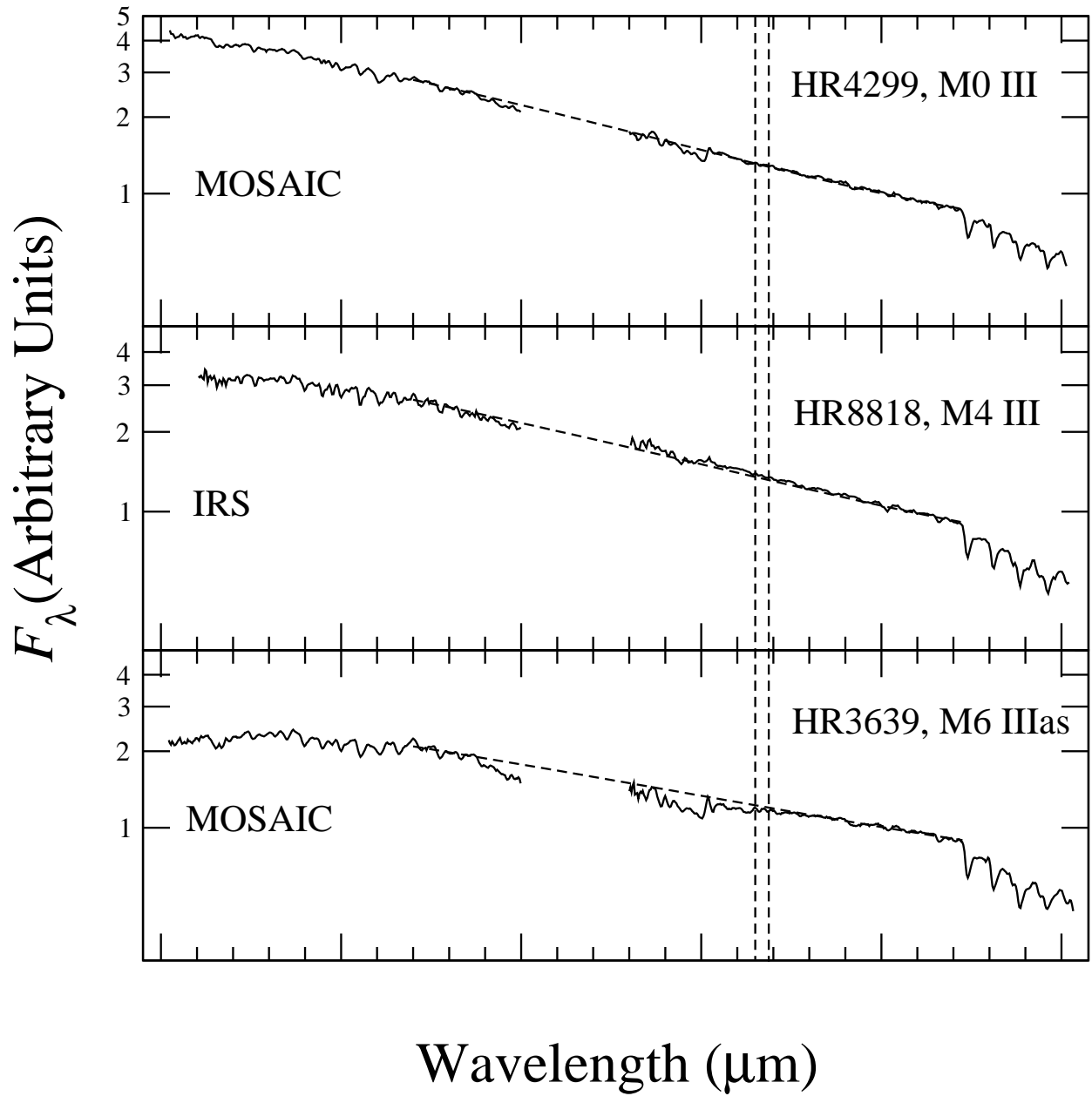} 

\caption{Same as Figure~\ref{sg} but for late--type
giants. HR3639 has similar CO strength compared to 3 Cet, but visibly
stronger \h2o \ (Figure~\ref{sg}). Y axis is scaled logarithmically.
\label{mg}}
\end{figure}

The tabulated \teff \ and computed CO indices for the comparison stars
(see Table~3) were used to derive a linear relationship for \teff \ vs
CO strength. The relationship (a least squares fit to the data) is
shown for both giants (\teff \ $=$ 4828.0 -- 77.5$\times$CO) and
supergiants (\teff \ $=$ 5138.7 -- 68.3$\times$CO) in
Figure~\ref{covteff}. 
The offset between the two luminosity classes is
due to the effects described in \citet{kh86} and BSD96.
Figures~\ref{sg} and \ref{mg} show
several supergiant and giant spectra from our sample.

The four points for LPVs shown in Figure~\ref{covteff} effectively
constitute a third relationship for \teff \ vs CO strength; however,
the CO strength vs \teff \ for LPVs does not correlate in the same way
as for giants and supergiants (increase in CO strength for lower
\teff). This is most likely due to the fact that there is such strong
\h2o \ absorption in these stars that the CO continuum band near 2.29
\mic \ is affected (depressed); see Figures~\ref{lpv} and
\ref{covteff}. The coolest LPVs have weaker CO. We discuss the determination
of \teff \ for the LPVs in the next section. 

\subsubsection{\h2o \ Index}

Given a measured CO strength, \teff \ can be determined using the
calibration from the comparison stars (Figure~\ref{covteff}), provided the
luminosity class is known. For purposes of determining \mbol, we take
``LPV'' as a distinct luminosity class because we assign different
bolometric corrections to giants (AGB, III), supergiants (I), and LPVs
(Miras, tip of the AGB).

The \h2o \ strengths for our comparison stars are generally small for
both the giants and supergiants because the giants are dominated by
earlier M spectral types (a consequence of our \mbol \ and \teff \
selection), and the supergiants do not exhibit strong \h2o \ as
described above.  \h2o \ strengths are given in Table~3 for the
comparison giants and supergiants. The quoted uncertainty is derived
from the scatter of all the giant and supergiant stars.  The LPV
comparison stars, on the other hand, exhibit strong \h2o \ absorption,
32 $\%$ on average for the four stars listed in Table~3.

\begin{figure}
\plotone{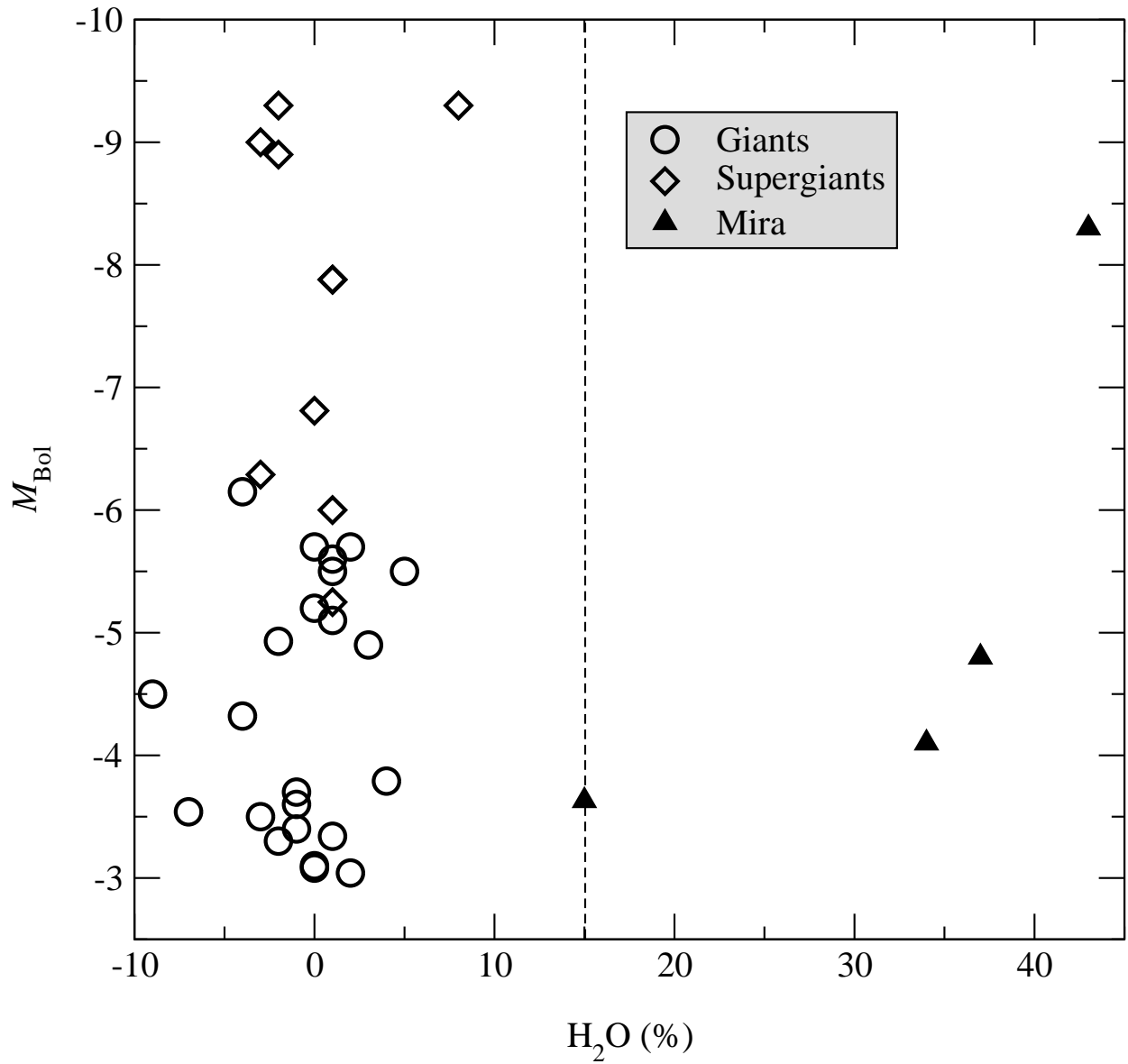}

\caption{\h2o \ absorption strength for comparison stars. Stars
with \h2o \ $\ge$ 15 $\%$ are known Miras (long period variables, or
LPVs).  All Galactic center stars with \h2o \ greater than 15 $\%$
were classified as LPV candidates with a correspondingly lower \teff;
see text.
\label{h20vmbol}}
\end{figure}

For purposes of classification, we will assume the \h2o \ strengths of
late M stars on the AGB lie between those for giants and LPVs. We take
the \h2o \ strength of R~Ser (\h2o \ $=$ 15 $\%$) to be the lower
bound for stars which are likely to be LPVs; see
Figure~\ref{h20vmbol}. This star is classified as a Mira (i.e. LPV)
but has less \h2o \ absorption than the other luminous LPVs in our
sample. Thus any star with \h2o \ $>$ 15 $\%$ in the GC will be
classified as an LPV candidate.  Table 3 shows that \teff \ depends on
the \h2o \ index for LPVs, in the sense that \h2o \ is stronger for
cooler stars.  We use this relationship (\teff \ $=$ 2893.0 - 8.8
$\times$ \h2o) to derive \teff \ for GC stars which are likely to be
LPVs based on their \h2o \ strength. LPVs are assigned an uncertainty
in \teff \ based on the full range of values for the comparison stars
($\pm$ 200 K).

For non--LPVs in the GC, the relations between CO and Teff for giants
and supergiants, illustrated in Figure~\ref{covteff}, are used to
derive \teff.  The distinction between giants and supergiants,
however, is not entirely straightforward.  The classifications were
initially made by eye, paying attention to the appearance or absence
of \h2o \ between the $H$ and $K$ bands for a given CO strength.
However, we then determined what quantitative values of CO, \h2o \,
and \mbol \ reproduced our classifications by eye.  We used \mbol \ to
make an initial distinction (see \S4.1 for the derivation of
\mbol): GC stars with \mbol \ $<$ $-$7.2 must be
supergiants (see discussion in BSD96), while GC stars with \mbol \ $>$
$-$4.9 (the faintest \mbol \ known for supergiants; see Table 1 and
BSD96) are likely to be giants. The GC stars with $-4.9 >$ \mbol \ $>
-7.2$ could be either giants or supergiants, based solely on \mbol.
                                                                                
\begin{figure}
\plotone{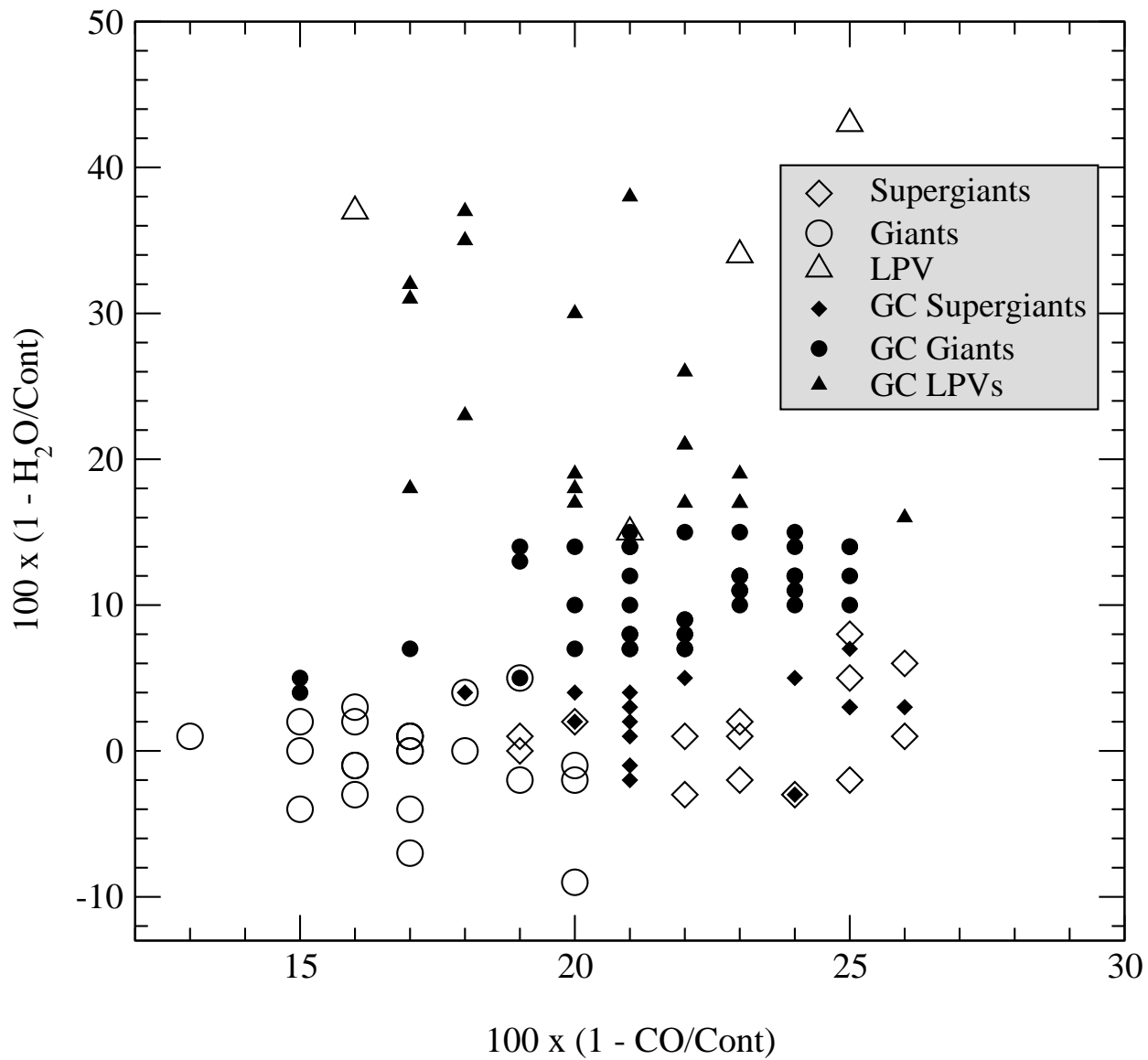}

\caption{\h2o \ strength vs CO strength for the Galactic Center
(GC) and comparison stars. The comparison stars are shown as {\it
open} symbols while GC stars are plotted with {\it filled}
symbols. For a given CO strength, GC stars classified as III have
larger \h2o \ than those classified as I. The GC III stars lie between
the warmer comparison stars and cooler comparison star LPVs. GC stars
with \h2o \ $>$ 15 $\%$ are classified as LPVs or LPV candidates (LPV?
in Table 6). \label{covh2o}}

\end{figure}

For the GC stars with $-4.9 >$ \mbol \ $> -7.2$, we use a combination
of CO and \h2o \ to distinguish giants from supergiants.  This is
shown in Figure~\ref{covh2o}.  Supergiants have CO $> 20 \%$, and a linear
relationship between \h2o \ and CO defining the upper envelope
of supergiants, such that \h2o \ $<$ $-5$ + (0.5 $\times$ CO), between CO
indices of 20 $\%$ and 26$\%$, for supergiants.  Giants are
all the remaining GC stars which have not previously been
classified as LPVs or supergiants by these techniques.
The bolometric correction (\S4.1) depends on assigned luminosity class, and
most of the giants with \mbol \ $>$ $-$4.9
could be more luminous than \mbol \ $=$ $-$4.9
if the luminosity
class was assumed originally to be supergiant instead. However, each case
would then result in a giant classification based on the linear
relation between CO and \h2o. Thus, the original assignments based on
the appearance of \h2o \ are consistent with the assignments based
on the measured CO, \h2o, and derived \mbol.
For stars in common with BSD96, we have arrived at the same
luminosity classes, and also for star VR5--7 of the Quintuplet cluster.
\citet{mgm94} classified as VR5--7 as a late--type M supergiant.

As an aside, we note that for two of the comparison stars (R~Dor and
EP~Aqr), we could not compute reliable \h2o \ indices owing to
possible low--order variations and/or slope changes in the
continuum. This may be due to the non--standard data taking procedures
which were used to keep these bright stars from saturating (see above)
although most of the stars appear to have normal spectra independent
of how they were obtained. These variations should not affect the
relatively narrow and closely spaced CO and associated continuum
measurements. R~Dor and EP~Aqr are the two coolest giants not
identified as Miras or LPVs in our sample, yet our spectra show no
evidence of \h2o \ absorption when fit for a continuum like the other
stars.

\subsection{Galactic Center Sample}

The GC stars are listed in Table~2. The complete GC sample was chosen
as all stars in the GC $K-$band luminosity function (KLF, taken here
to be the {\it dereddened} luminosity function) as derived by BDS96
which have $K_{\circ}$ $\le$ 7.0 (where $K_{\circ}$ is the
dereddened $K$ magnitude).  We have revised several stars to
$K_{\circ}$ magnitudes fainter than 7.0 based on new, unpublished
$H-$band images used to determine $A_K$ (see the footnote to Tables~2
and 4). With these adjustments, there are 136 stars in the complete
sample. However, 22 stars in this original list were not observed
because of severe crowding with neighboring stars, and we were unable
to observe an additional 24 stars because of cloudy weather at the
telescope. Eleven stars in the list are known emission--line stars or
have featureless continua (BDS96, Tables~2 and 4). The featureless
stars are apparently young massive stars which are embedded in ionized
gas filaments in the central pc \citep{tan03}. We thus take them to be
part of the youngest burst of star formation in the GC.  All 136 stars
are listed in Table~2. Observation dates and instruments are given for
the stars for which we obtained new spectra.

Table~5 indicates the level of completeness in the spectroscopic
sample as a function of luminosity.  In what follows, we assume the
stars in our complete list which we were not able to observe are
distributed in the Hertzsprung--Russell Diagram (HRD) like the cool
stars we have observed. This is a good assumption since previously
known emission--line objects (apart from those in Table~5) are fainter
than $K_{\circ}=$ 7.0, and all of the stars we observed in this work
from the complete sample were cool stars.

\begin{figure}
\plotone{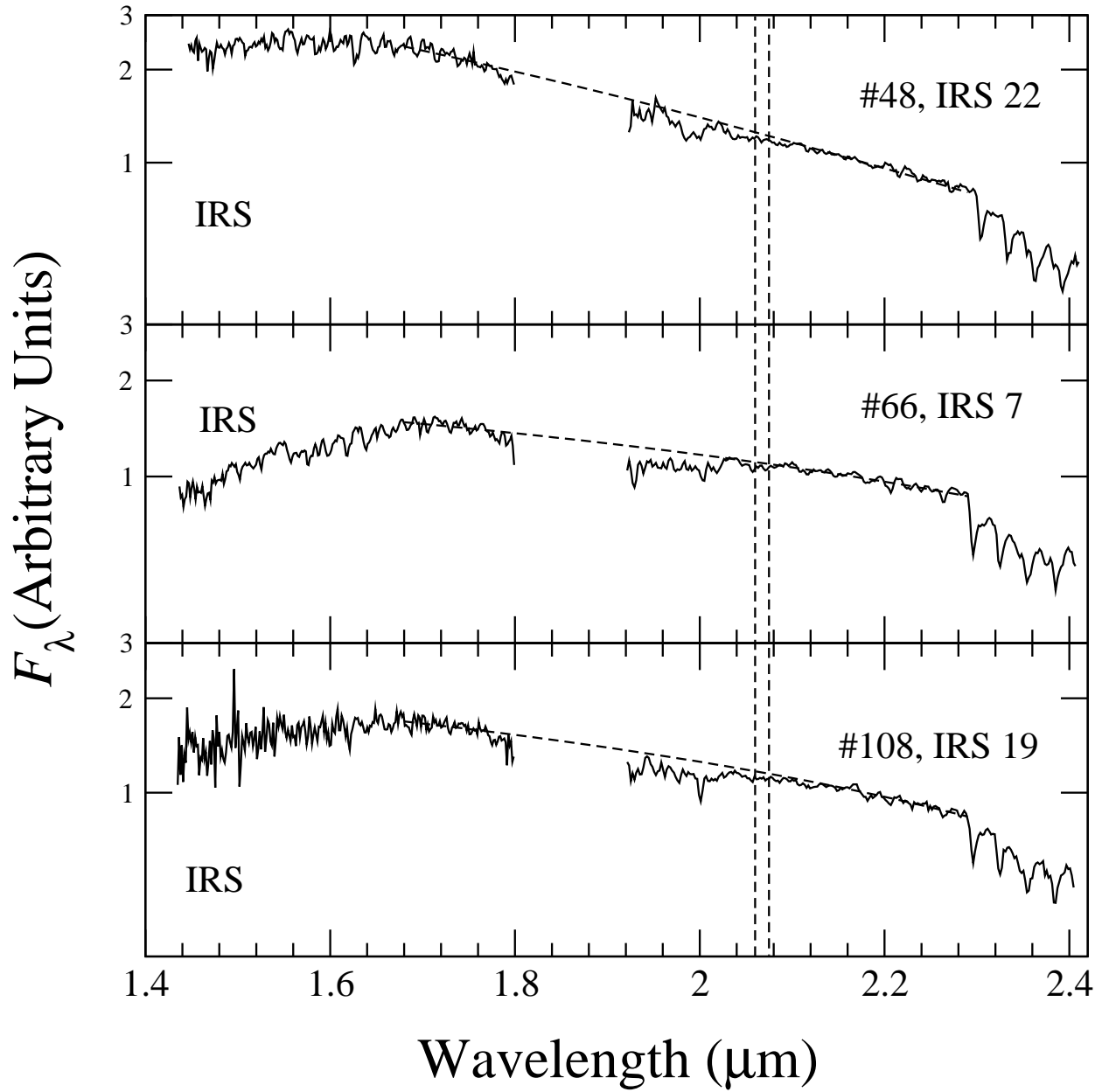}

\caption{Example spectra of stars classified as supergiants in
the Galactic center. These stars were analyzed at high spectral
resolution by \citet{ram00} and \citet{csb00}. The {\it dashed} curves
are the fits to the continua used to measure \h2o \ at the position of
the vertical {\it dashed} lines (see text and Figure~\ref{lpv}). Y
axis is scaled logarithmically.
\label{specA}}
\end{figure}

\begin{figure}
\plotone{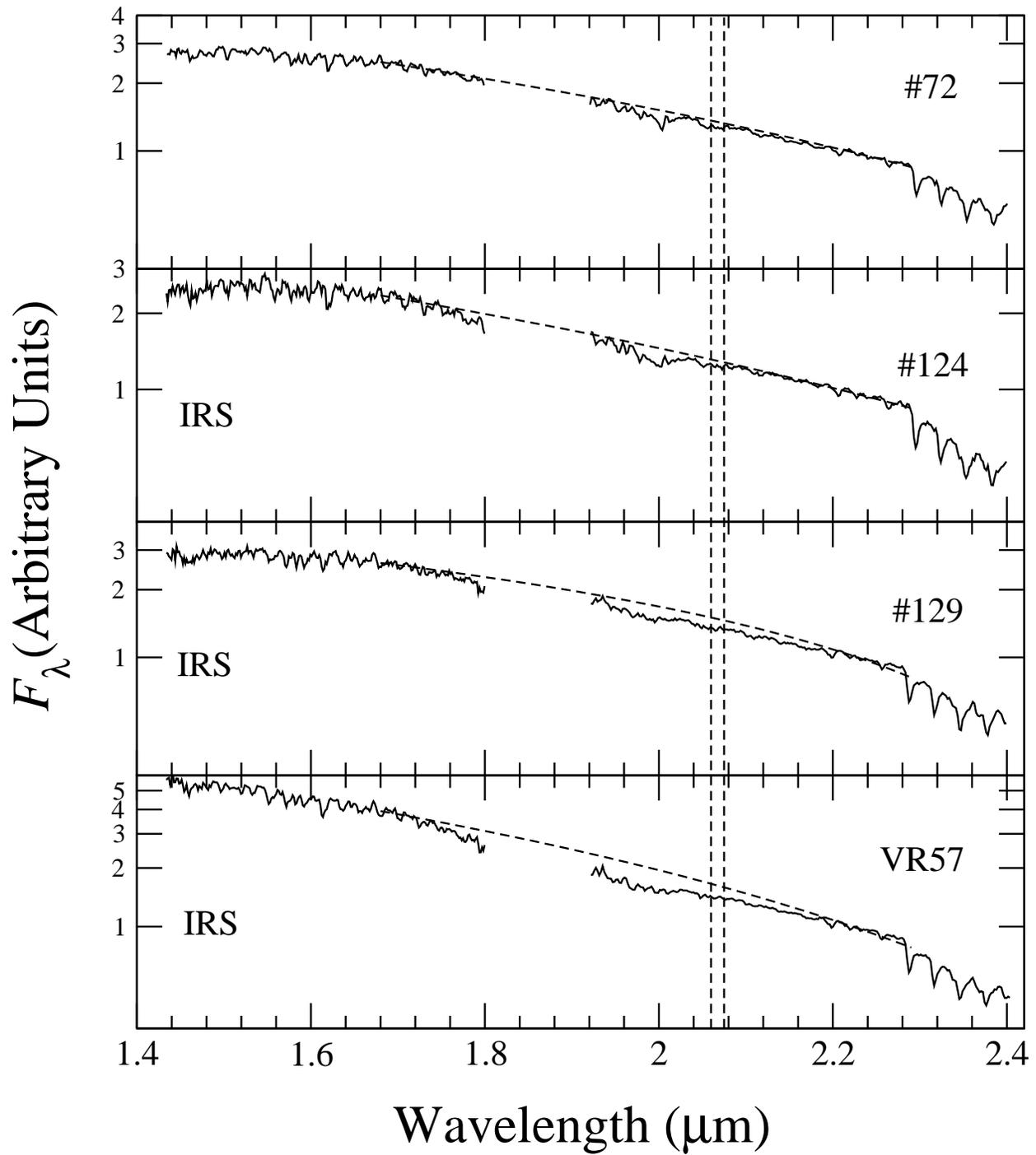} 

\caption{Same as Figure~\ref{specA}. Star 72 is classified as a
giant; all others are supergiants. These stars were analyzed at high
spectral resolution by \citet{ram00}.
\label{specB}}
\end{figure}

\begin{figure}
\plotone{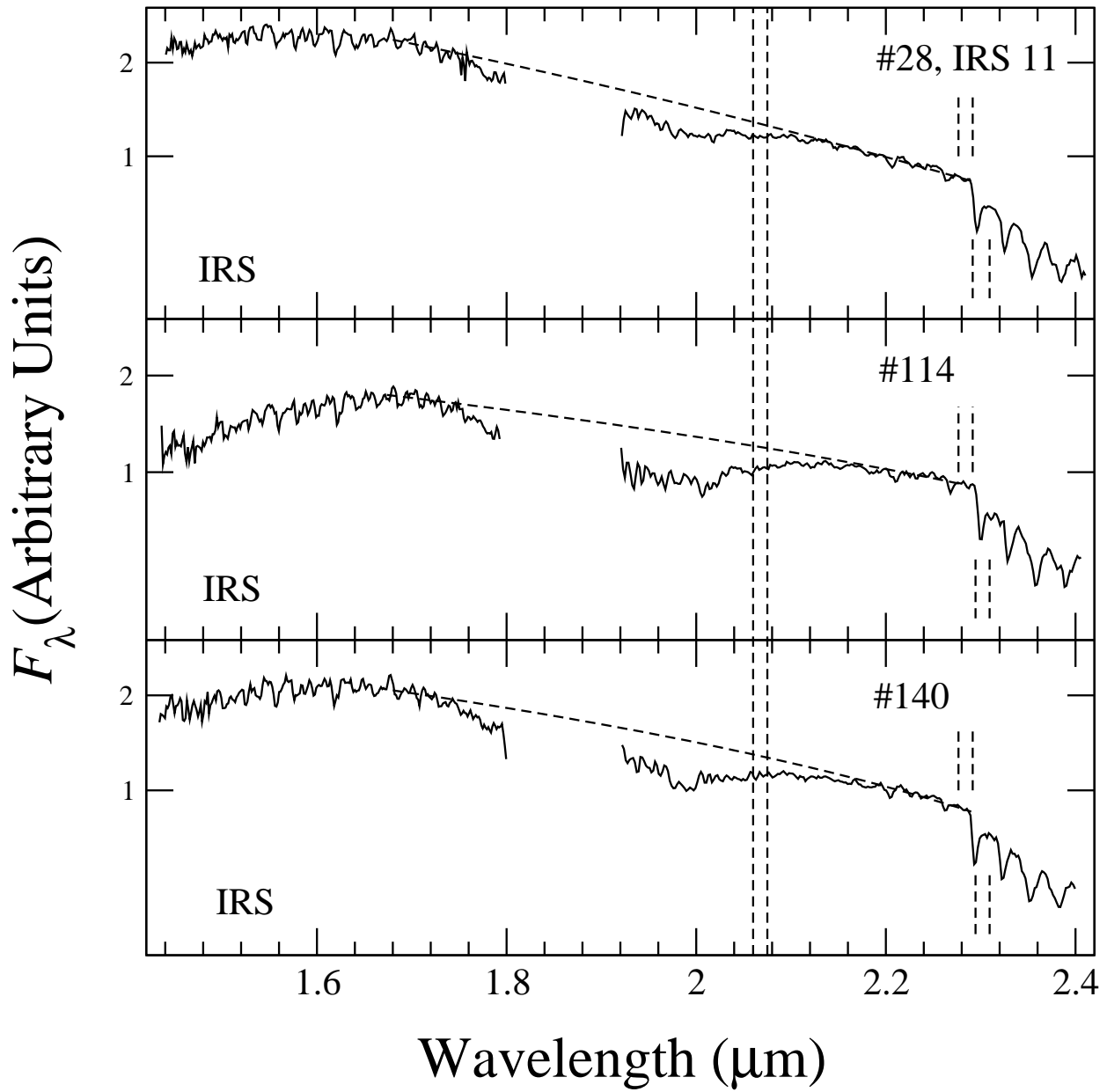} 

\caption{Example spectra of stars classified as asymptotic giant
branch (referred to in the text as AGB, giant, or III stars) stars in
the Galactic center. Stars \#28, \#114, and \#140 were analyzed at
high spectral resolution by \citet{ram00}. The {\it dashed} curves are
the fits to the continua used to measure \h2o \ at the position of the
vertical {\it dashed} lines (see text and Figure~\ref{lpv}).  Y axis
is scaled logarithmically.
\label{spec2}}
\end{figure}

\begin{figure}
\plotone{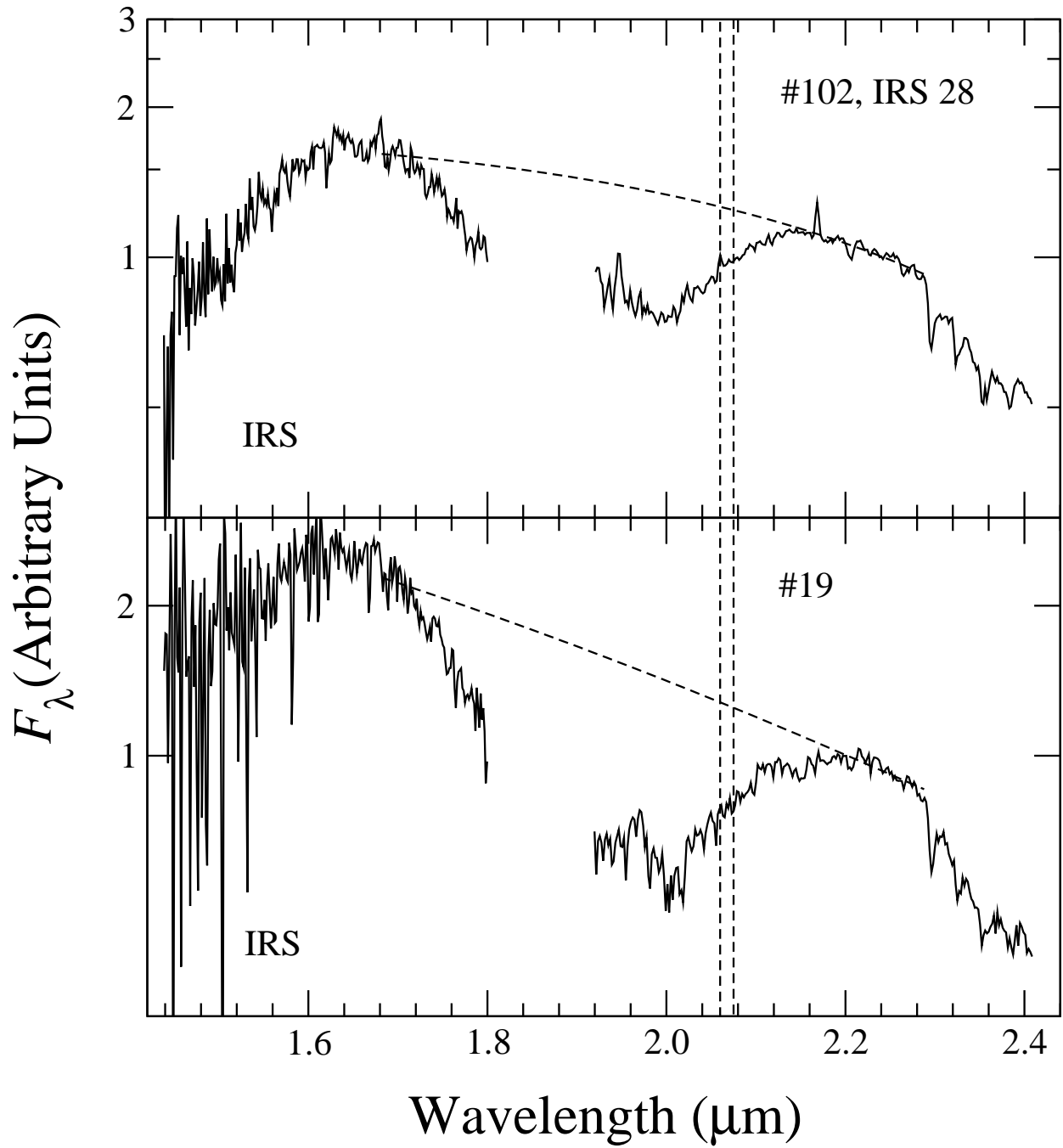} 
\caption{Stars classified as long period
variables (LPV) or candidate LPVs (LPV? in Table~6) in the Galactic
center.  IRS 28 is a known
photometric variable; see text. The emission--line near 2.17 \mic \
evident in the spectrum of IRS~28 is likely due to incomplete
subtraction of the local nebular background. The {\it dashed} curves
are the fits to the continua used to measure \h2o \ at the position of
the vertical {\it dashed} lines (see text and Figure~\ref{lpv}). Y
axis is scaled logarithmically.
\label{spec3}}
\end{figure}

Example spectra of GC supergiant, AGB, and LPV stars are shown in
Figures~\ref{specA}$+$\ref{specB}, \ref{spec2}, and \ref{spec3}
respectively. The complete sample of cool stars observed at high
spectral resolution by \citet{csb00} and \citet{ram00} is shown in
these figures, including all stars for which the values of \teff \ in
this paper were used in the abundance analysis of \citet{ram00}.

Table~6 lists the measured CO and \h2o \ strengths for each GC star.
Four stars in our list are taken from BSD96 and have only $K$ band
spectra (IRS~1NE, IRS~1SE, OSUC3, and OSUC4). Classifications for
three other stars (\#'s 170, 173, and 214) were made based only on
their $K-$band spectra because use of the $H-$band spectra (which was
much fainter) in these cases resulted in poor fits for the continuum
and hence \h2o \ indices which were clearly in error.

The GC star luminosity class is given in Table~6 for each GC star.  As
noted above, any GC star with \h2o \ $>$ 15 $\%$ is classified as an
LPV (or candidate, LPV? in Table~6). We retained the classification of
LPV for stars IRS~9, IRS~12N, IRS~24, and IRS~28 of BSD96. These are
known photometric variables (Haller et al. 1992, Tamura et al. 1996,
BDS96, Ott et al. 1999). For LPVs and LPV?s, the average \h2o \ was 23
$\%$ $\pm$ 7 $\%$. Twenty (20) GC stars were classified as LPVs out of
the spectroscopic sample of 79.

\begin{figure}
\plotone{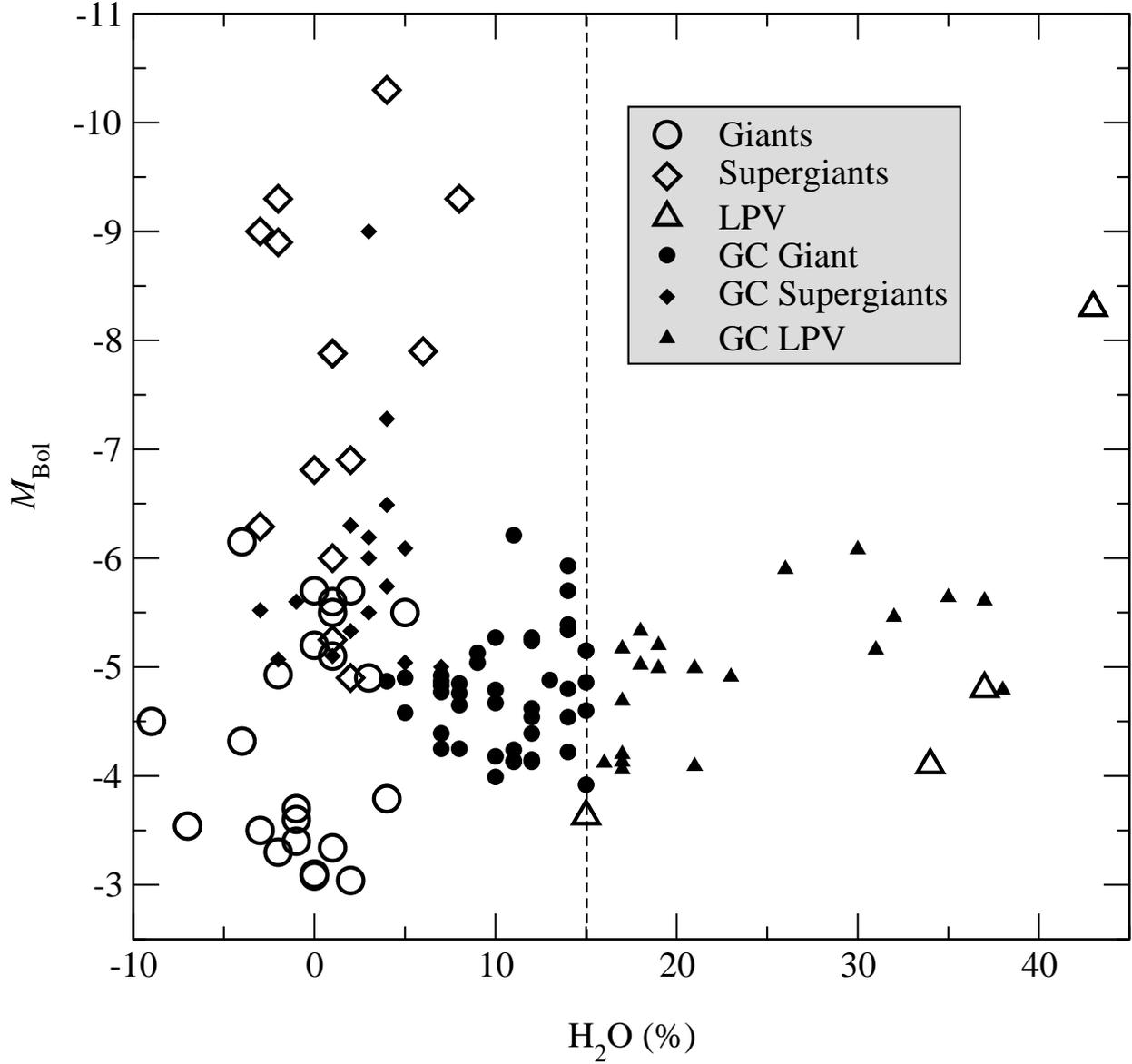} 

\caption{Comparison and Galactic Center (GC) star \mbol \
vs. \h2o \ strength plot. The comparison stars are shown as {\it open}
symbols while GC stars are plotted with {\it filled} symbols. The
assigned GC luminosity classes are consistent with the comparison star
luminosity ranges. The plot shows we have observed later type giants
in the GC than are represented in the warmer comparison III sample,
and these GC IIIs lie between the comparison IIIs and LPVs, not along
the comparison I track which runs vertically in this diagram (see text
for details, and also Figure~\ref{covh2o}).
\label{h2ovmbol2}}
\end{figure}

GC stars with lower \h2o \ were binned into III and I classes based on
the appearance (by eye) of the overall absorption between the $H$ and
$K-$bands and also the measured quantities for CO and \h2o \ as
described n\S3.2.2: stars with strong CO compared to the comparison I
stars can have slightly stronger \h2o \ and still be classified as I,
while stars with less CO must have very low measured \h2o \ to be
classified as I. The GC luminosity class assignments (based only on
the \h2o \ and CO absorption) are consistent with the comparison star
luminosities.  If we plot the GC and comparison star \h2o \ vs \mbol,
we see the GC stars fall along two broad tracks delineated by the
comparison stars (Figure~\ref{h2ovmbol2}). To the lower left in this
diagram are the warm comparison star III's. Going vertically at low
\h2o \ are the supergiants. The comparison star LPV's run generally to
the right to larger \h2o \ absorption and to gradually higher
luminosity. The GC I's follow on or near the comparison I track, while
the cooler GC III's essentially fill the region between the warmer
comparison III's and the cooler comparison LPVs as is expected for
cooler III's high on the AGB. Several GC stars exist where these broad
tracks overlap. In each case, the GC star \mbol, is consistent with
the comparison luminosities for the corresponding class.  The average
\h2o \ strength for GC stars classified as giants (III in Table~6) was
10 $\%$ $\pm$ 3 $\%$. For supergiants, the average was 3 $\%$ $\pm$ 2
$\%$.  The GC stars are also well separated between I and III classes
in the \h2o \ vs. CO plot (Figure~\ref{covh2o} and consistent with the
positions of the comparison stars.

In summary, we used criteria on the appearance and measured strength
of \h2o \ and measured CO strength to assign all GC luminosity
classes.  We then showed that the GC assignments, when quantitatively
based on CO, \h2o, and \mbol \ reproduced the classifications by eye
and are well matched to the parameters of the comparison stars. The GC
giants lie at or below the comparison giant luminosities while
supergiants always lie above the minimum comparison supergiant
\mbol. As noted in \S3.2.2, the comparison star sample does not
include giants later than type M6 as a consequence of our selection
based on \mbol \ and \teff: such stars are rare in the solar
neighborhood.  Nevertheless, the criteria used here for CO and \h2o \
indicates that such such stars exist in larger numbers in the GC as
would be expected in this dense stellar environment.

\section{RESULTS: DETERMINATION OF THE STAR FORMATION HISTORY}

\subsection{The Hertzsprung--Russell Diagram}

The spectral indices developed in the preceding section allow us to
derive bolometric magnitudes and \teff \ for the GC stars. \teff \
follows directly from the measured CO index once the luminosity class
is chosen (see \S2.3). \teff \ values are given in Table~6.  The
uncertainty in \teff \ is derived by propagating the uncertainty in
the measured CO strength through the CO vs. \teff \ relation.  This
gives an average uncertainty, for all giants and supergiants in the
sample, of 184 K.  For the 11 stars in common with the sample given by
BSD96 (not including the LPVs), we find $\Delta$\teff \ $=$ 32 K $\pm$
156 K.  In addition, our value of \teff \ for VR5--7 differs from that
calculated by \citet{ram00} by only 74~K. The \mbol \ for a given GC
star follows from the intrinsic $K_{\circ}$ magnitude given in
Table~2, the distance modulus, and a bolometric correction to the
$K_{\circ}$ magnitude.  The uncertainty in \mbol \ is the sum in
quadrature of the photometric uncertainty (BSD96) and an 0.4 mag
uncertainty due to the uncertainty in the interstellar extinction law
\citep{m90}.  The distance modulus is taken as 14.52 (8 kpc, Reid
1993).

The bolometric correction (BC$_K$) is derived from the literature for
the different luminosity classes given in Table~6.  For supergiants
and LPVs (Table~6), the BC$_K$ is the same as given by BSD96: BC$_K =
$2.6 for supergiants and 3.2 for LPVs. For the giants, we improve on
the work of BSD96 by considering a BC$_K$ which is a function of
\teff. Using the BC$_K$ as a function of $J - K$ given by
\citet{fw87}, the mean $J - K$ of giants as a function of spectral
type from \citet{fpam78}, the spectral type vs \teff \ from
\citet{ram00} and \citet{dy98}, and the \teff \ given in Table~6, we
derive BC$_K$ as a linear function of \teff \ (BC$_K$ $=$ 
2.6 $-$ [\teff \ $-$ 3800]/1500). These BC$_K$ range from
2.8 to 3.2 for the warmest and coolest GC giants.

A second correction was also applied to the GC stellar \mbol \ values
which relates to the $A_K$ for each star. BDS96 used mean $(J -
H)_{\circ}$ and $(H - K)_{\circ}$ colors to estimate the $A_K$ for
each GC star. As discussed by them, this will lead to systematically
too high $A_K$ for stars which are intrinsically redder than these
colors, and the opposite will be true for stars intrinsically
bluer. The individual spectrum for each star now allows us to improve
upon the corresponding $A_K$ estimate. Using the same color and \teff
\ data described in the preceding paragraph, we estimated a correction
(linear with \teff) to the intrinsic colors, and hence to $A_K$ for
each star. This correction varied from 0.0 to 0.5 mag depending on
\teff \ (see Table~6) and was applied in the sense which makes $A_K$
less for each star ($\Delta A_K$ $=$ $-$0.11 $-$ [\teff \ $-$
3800]/1730).  

\begin{figure}
\plotone{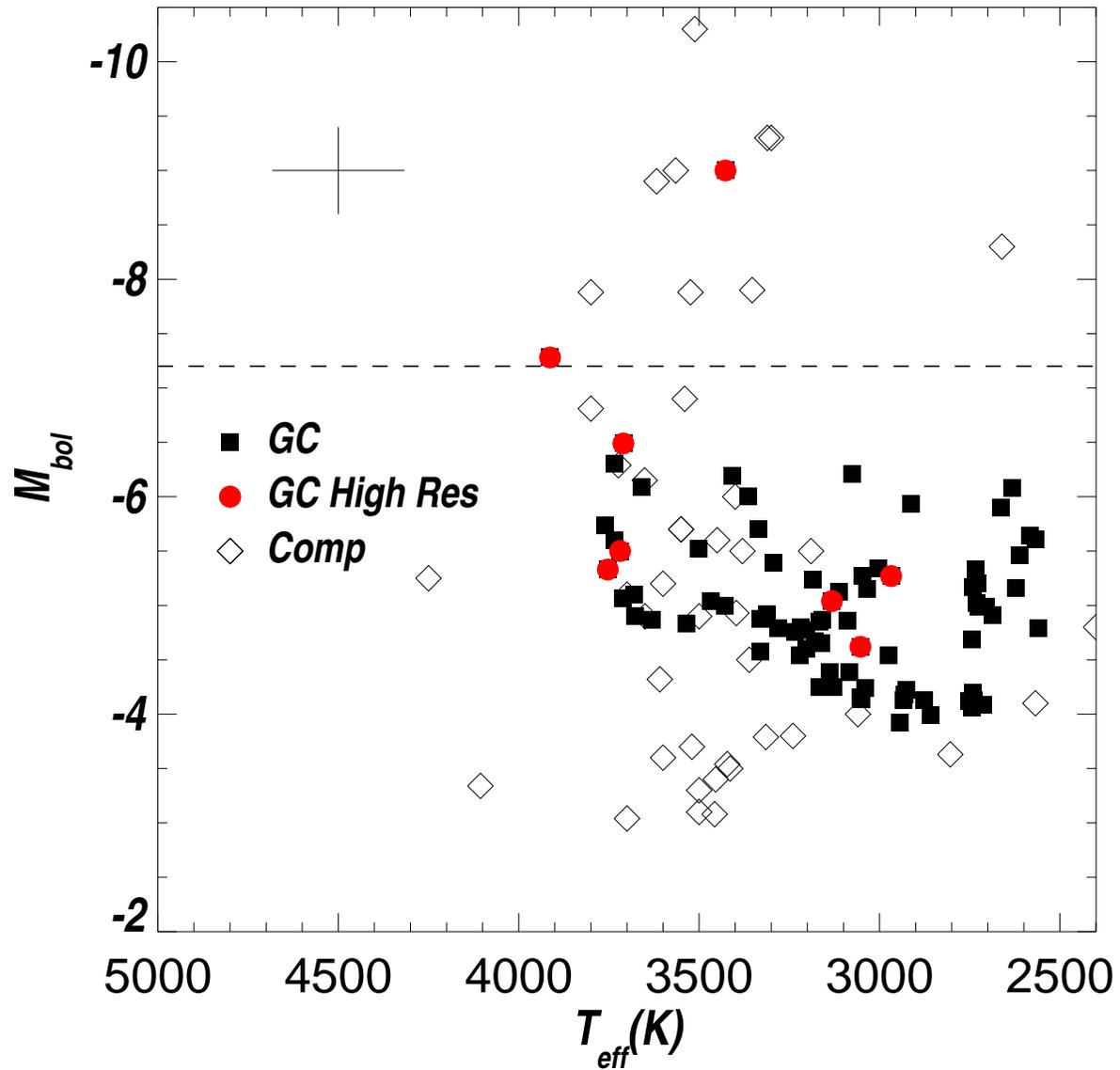} 

\caption{Hertzsprung--Russell diagram for the Galactic center (GC)
stars (shown as {\it closed} squares with typical uncertainty given by
the error bar in the upper left corner) and comparison
stars (Comp, plotted as {\it open} diamonds).  The GC stars
analyzed at high spectral resolution by \citet{csb00} and
\citet{ram00} are plotted as {\it filled} circles. \label{hrd}}
\end{figure}

Figure~\ref{hrd} shows the HRD for all the GC stars and comparison
stars listed in Tables 3 and 6.  This figure illustrates that the
comparison stars span the same range in \mbol \ and \teff \ as the GC
stars do.

\begin{figure}
\plotone{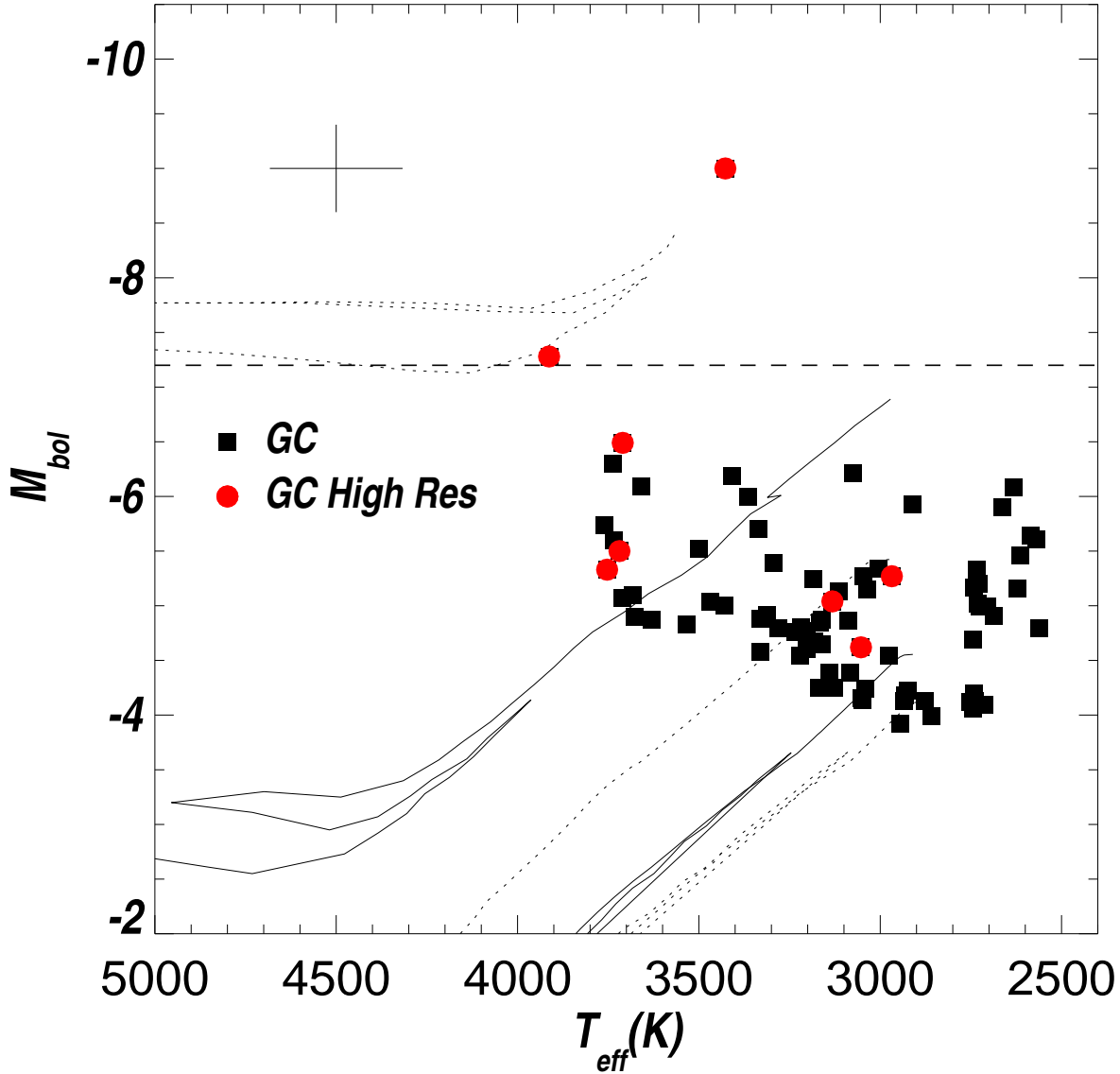}

\caption{Galactic center (GC) stars plotted as in Figure~\ref{hrd}, 
but with [Fe/H] $=$ 0.0 isochrones plotted as well.
The isochrones are from \citet{bert94} for
age $<$ 100 Myr and \citet{gir00} otherwise. Isochrones are plotted
for ages of 10 Myr, 100 Myr, 1 Gyr, 5 Gyr,
and 12 Gyr.  The models have [Fe/H] $=$ 0.0 for all ages, and these
appear to better represent the data than models with lower
metallicity at older ages ([Fe/H] $=$ $-$0.2); see Figure~12$c$ and text.
Neither set of isochrones reaches the coolest stars (long period variable
candidates with \teff \ $<$ 2800 K), but the [Fe/H]
$=$ 0.0 isochrones extend to cooler temperatures and thus fit more
Galactic stars than the [Fe/H] $= -0.2$ isochrones.
Comparison to the isochrones shows that all the GC stars classified as
giants (Table~6, III) are AGB stars; they are too luminous to be first
ascent giants. This is a consequence of the selection criteria. The
horizontal line segment at $M_{\rm bol} = - 7.2$ in each panel
indicates the approximate observed luminosity above which only
supergiants lie \citep{bsd96}.
\label{hrd0}} 
\end{figure}

Figure~\ref{hrd0} shows the HRD for all the GC stars listed in Table
6, with isochrones from \citet{bert94} and \citet{gir00} overplotted.
These isochrones vary in age from 10 Myr to 12 Gyr, with [Fe/H] $=$
0.0 for all ages.  Figure~\ref{hrd0} shows that our GC sample spans a
wide range in age.  As can be seen in the Figure, all of the GC giants
(those labeled "III" in Table 6) are AGB stars.  They are too luminous
to be first ascent giants, which is a consequence of our selection
criteria.

\begin{figure}
\plotone{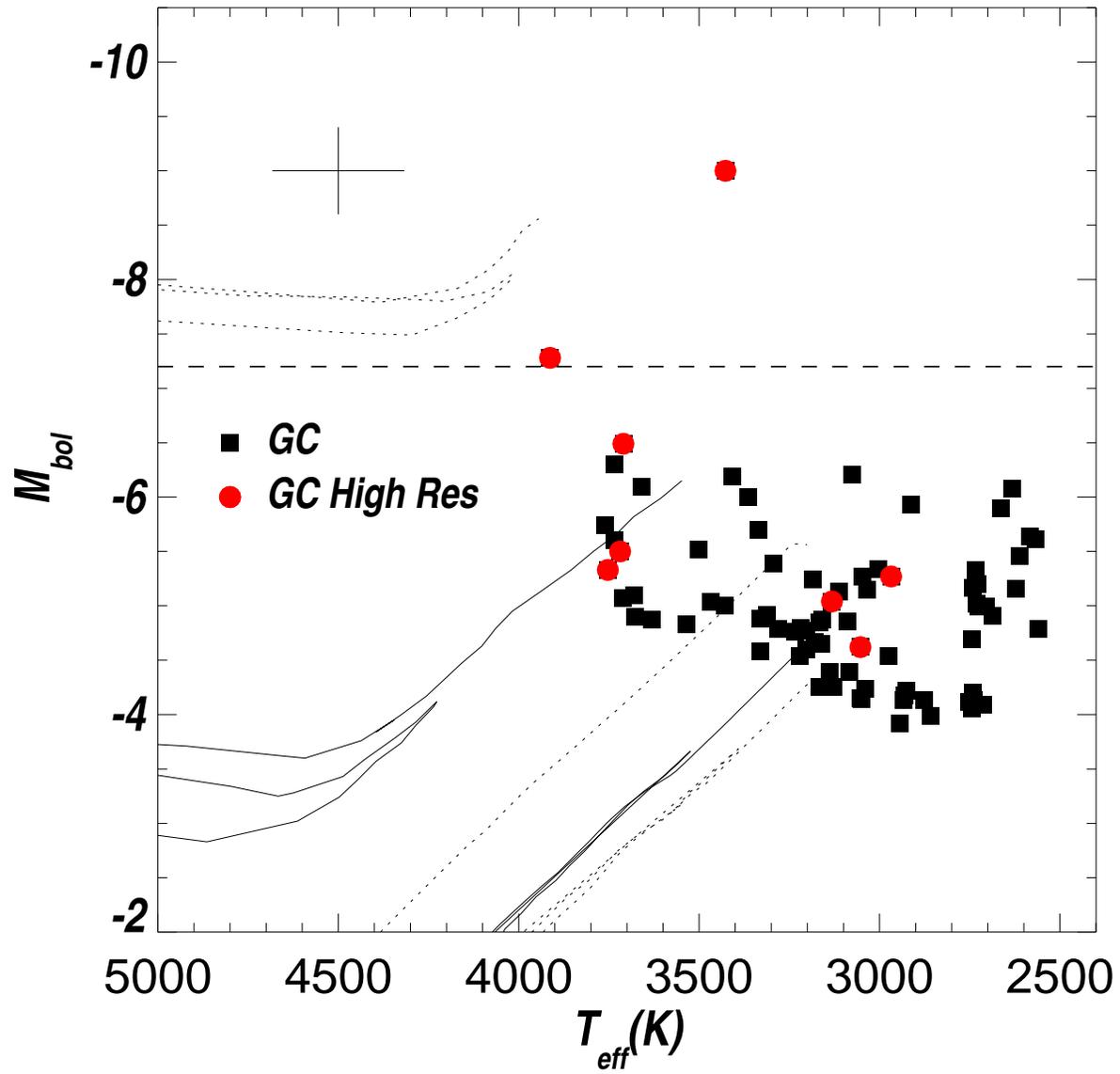}
\caption{Same as for Figure~12$b$, but with [Fe/H] $=$ $-$0.2 
isochrones plotted.
\label{hrd2}}
\end{figure}

Figure~\ref{hrd2} again shows the HRD for the GC, but this time
overplotted with \citet{bert94} and \citet{gir00} isochrones having
[Fe/H] $=$ $-$0.2 for all ages. The Figure shows that lower
metallicity isochrones do not extend to cool enough \teff \ to match
the GC HRD.

To give a general feel for the SFH represented by Figure~\ref{hrd0},
Figure~\ref{testhrd} shows the HRD against a simulation representing a
constant SFH.  The observed points appear to span the model parameter
space.  The relatively few model stars at high and low luminosities
suggests that higher star formation rates in both the distant and
recent past are needed to fit the data.  There is an intermediate age
component at \teff \ $=$ 3300 K, \mbol \ $=$ $-$5.0 whose position and
extent matches the model well. The tightness of this feature suggests
that our errors may be over estimated relative to the internal scatter
in the points. This is particularly true for the errors in \mbol \
which are dominated by the {\it systematic} uncertainty in the
interstellar extinction law.

One exception in the model coverage appears to be the coolest AGB
stars. Both the \citet{bert94} and \citet{gir00} models fail to reach
the coolest observed \teff. This is true of both the GC and comparison
stars. Thus, only stars with \teff \ $>$ 2800 K (allowing for the
errors in \teff \ for the observed stars) were directly included in
the basic SFH calculations discussed below. The models attempt to
follow evolution along the AGB in a simplified way \citep{gir00}
giving a typical or average locus in the HRD, but real AGB stars
pulsate with periods of 100's of days. These pulsations result in
excursions in the HRD of 500--1000 degrees \citep{lan02}.  As
discussed by \citet{lan02}, the effect of pulsations is thus to widen
the AGB. For stars which experience excursions within the model
temperature range (\teff \ $>$ 2800 K), the pulsations will be
randomly phased, so that differences between the observed location in
the HRD and the actual isochrone to which a star would otherwise be
associated with are canceled out.  For stars which are cooler than the
models, we assume they should be associated with an isochrone inside
the model HRD space. Assuming that these stars represent the same
fraction of initial mass independent of which isochrone they are
really associated with, we simply scale the total star formation by
their number. In this case, there are 20 such cool stars (out of 78
used to calculate the SFH), thus we will take the total star formation
rates to be 1.3 times the amount given by our fit results.
 
We also did not include IRS~7. This star belongs to the youngest
nuclear star burst which is accounted for by the \citet{krabbe95}
model.

BSD96 estimated ages for a number of the coolest, most luminous stars
which are also shown in Figures~\ref{hrd}, \ref{hrd0}, and \ref{hrd2}.
The effect of a reduced $A_K$ for these stars (compared to BSD96) has
lowered their luminosity. This tends to increase the estimated age.
In particular, for IRS 9, 12N, 23, 24, and 28 (see Table~6), BSD96
estimated the mass for these stars from their luminosity. This
resulted in correspondingly young ages (100--200 Myr) from model
isochrones. The reduced luminosity determined in the present work (due
to the decrease in $A_K$), and comparison to different model
isochrones as adopted herein (Figure~\ref{hrd0}) suggests a somewhat
older age for these stars (roughly 500 Myr -- a few Gyr). Given the
preceding discussion, it is not possible to estimate the age of any
particular star to great precision.

\subsection{Star Formation History Calculation}

We have used the results of Figure~\ref{hrd} to derive the SFH implied
by these observations of the GC cool stars. The calculation was
carried out using Olsen's (1999) implementation of the method
described by Dolphin (1997), with some modifications.  In brief, we
constructed a set of models describing the expected distribution of
stars in the H--R diagram within specified age bins, assuming a
particular metallicity, slope of the initial mass function (IMF), and
constant star formation rate (1 M$_\odot {\rm yr}^{-1}$) within the
bin and accounting for observational errors and incompleteness.  We
chose the best model SFH for the GC by fitting the observed data to a
linear combination of the star formation within these bins.  This fit
was determined through the maximum likelihood analysis described
below.

\subsubsection{Model Parameters}

Two choices of sets of age bins and metallicities were used. Models
had either four age bins (Model A, 10--100 Myr, 100 Myr -- 1 Gyr, 1--5
Gyr, and 5--12 Gyr) or three age bins (Model 1, 10--50 Myr, 50 Myr --
3 Gyr, and 3--12 Gyr). For both sets of age bins, models were run with
all stars at solar [Fe/H] (Models A with four age bins and Model 1
with three age bins), and then again with solar [Fe/H] for the younger
stars and [Fe/H] $= -0.2$ in the oldest bin (Model B with 4 age bins,
Model 2 with three age bins). To explore the effect of the IMF on the
derived SFH, we also computed models with a power--law slope
$\alpha=-2.35$ \citep{s55} and with a slightly flatter slope
$\alpha=-2$ for stars with masses $>1 {\rm M}_\odot$.  The matrix of
models is given in Table 7.

The SFH is also constrained by the amount of mass inferred from
dynamical models in the GC. Recent models have been computed by
\citet{msbh89}, \citet{smbh90}, \citet{hall96}, \citet{genzel96},
\citet*{saha96}, \citet{ghez98}, and \citet{gen00}.  The most detailed
models, which include proper motion and radial velocities
\citep{ghez98,gen00} require a concentrated mass of approximately
3~$\times$ 10$^{6}$ M$_{\sun}$ (presumably a super massive black hole)
which dominates the distribution within less than 1 pc.  These models
also predict approximately 6 $\times$ 10$^{6}$ M$_{\sun}$ total mass
enclosed within a radius of 2 pc (for $R_\circ$ $=$ 8 kpc).  Finally,
new observations of single stars orbiting the black hole
\citep{scho02,ghez03} require slightly higher black hole masses of
$\approx$ 4 $\pm$ 1 $\times$10$^6$ M$_\odot$.

If the black hole itself was not built up from stars and stellar
remnants in the formation of the nucleus, this leaves roughly 1--3
$\times$ 10$^{6}$ M$_{\sun}$ in distributed mass which arises from the
luminous stellar population and the accumulated stellar remnants
integrated over the life time of the nucleus.

We built in the constraint on dynamical mass by limiting the low mass
end of the IMF. The constraint is taken to be: black hole mass $+$
stellar mass formed $-$ mass loss from stellar winds $=$ total
dynamical mass.  We discuss the effects of mass loss on the present
day mass below in \S5.  This is the simplest choice since our HRD is
not sensitive to stars below about 1 M$_\odot$. We discuss a possible
cause for this low mass cutoff below in \S5. We began with the same
IMF as \citet[and including their corrections]{meg00}, with masses
spanning the range 0.08--120 M$_{\sun}$. This IMF is based on
Salpeter's mass function for the more massive stars \citep{s55};
i.e. $dN/dM \propto m^{-\alpha}$ with $\alpha = -$2.35. For lower
masses, the IMF is flatter, as determined from bulge star counts at
6\deg \ projected distance from the GC \citep[exponent $-$2.0 for 1
M$_{\sun}$ $>$ M $>$ 0.7 M$_{\sun}$ and $-$1.65 for M $<$ 0.7
M$_{\sun}$]{z00}. For the lowest masses, there is a correction to the
exponent due to binaries \citep{meg00}.  In the end, we found that
cutting off the mass function at 0.7 M$_{\sun}$ resulted in models
which are consistent with the present day dynamical mass.

\subsubsection{Model Calculations}

Within each age bin, we calculated the distribution of stars over a
fine grid in $T_{\rm eff}$ and $M_{\rm bol}$ directly from 20
isochrones spaced linearly in age.  We used the \citet{gir00}
isochrones to construct the models with ages $\ge$ 63 Myr and the
\citet{bert94} isochrones for younger models; we interpolated the
isochrones in age and metallicity following the procedure described in
\citet{olsen99}.  We then convolved the grid with a 2--D Gaussian
kernel having a shape representing the typical errors in $T_{\rm eff}$
and $M_{\rm bol}$ and multiplied the grid with a surface representing
our estimate of the completeness as a function of $T_{\rm eff}$ and
$M_{\rm bol}$ (i.e., the models were transformed to the observational
plane).

After gridding the observed H--R diagram to the same resolution as the
models and selecting an area to exclude the likely Mira variable
stars, we searched for the linear combination of models producing the
highest likelihood of describing the observed distribution of stars.
This search was done by using the {\it Numerical Recipes} routine {\it
amoeba} (Press et al.\ 1992) to minimize the Poisson maximum
likelihood parameter $\chi^2_\lambda = 2\sum_{i}m_i-n_i+n_i{\rm
ln}\frac{n_i}{m_i}$ (e.g. Mighell 1999), where $m_i$ is the number of
stars predicted by the model in the $i$th bin of the H--R diagram and
$n_i$ is the number of observed stars in the bin. The virtues of this
parameter are discussed extensively by \citet{dol02}--the most
important of which is that it minimizes to the correct solution when
presented with a dataset that sparsely samples the range of possible
measurements (i.e. the Poisson regime), which the standard $\chi^2$
does not.

\subsubsection{Uncertainties and Goodness of Fit}

The size of the errors in $M_{\rm bol}$ and $T_{\rm eff}$, the size of
our dataset, and the fact that we are studying only the most luminous
members of the GC population impose some limitations on our ability to
discriminate model SFHs.  First, we selected the age bins so as to
roughly divide the H--R diagram into equal areas and to distinctly
separate the footprints of adjacent age bins, given our errors.
Because the isochrones bunch together at larger ages, the age bins
necessarily grow wider with age, with a corresponding decrease in age
resolution in the derived star formation histories.  Second, we
specified the age--metallicity relation in advance.  While the maximum
likelihood procedure described above in principle has the ability to
{\em derive} the metallicity distribution and age-metallicity relation
directly from the data, the degeneracy between age and metallicity in
this region of the H--R diagram is severe.  This difficulty is
compounded by our sizable errors in \mbol \ and \teff \ and the
relatively small size of our sample.  Finally as mentioned above, we
assumed the IMF, since our data do not span a large enough range in
mass at a given age to allow it to be a free parameter.

For each model in Table~7, we calculated the uncertainties in the
derived star formation rates through Monte Carlo simulations.  We
produced 100 Monte Carlo samples, each containing 59 stars, by drawing
randomly from the {\em observed} dataset while allowing any particular
star to be drawn any number of times (a technique referred to as
``bootstrapping'').  For each sample, we then derived the star
formation history just as was done for the original dataset.  The
uncertainty in a given star formation rate ($\sigma_{\rm SFR}$)
reported in Table 7 is the 1$-\sigma$ standard deviation of the
distribution of star formation rates in the corresponding age bin for
the 100 Monte Carlo samples.

To address the separate question of whether the data are a likely
representation of the models listed in Table 7, we ran a different set
of Monte Carlo simulations.  For these simulations, we produced 10000
samples, each containing 59 stars drawn randomly from the {\em fitted}
models ({\it not} from the {\it data}).  We then assembled the
distribution of minimum values of the $\chi^2_\lambda$ parameter by
re--fitting the model to each Monte Carlo sample.  The column labeled
$P_\lambda$ in Table 7 shows the percentage of runs that
had higher $\chi_\lambda^2$ when fitting the Monte Carlo sample to the
model than that obtained by a fit of the data to the model.
Thus, small values of $P_{\lambda}$
represent poorer fits of the models. This is so because Monte carlo
samples drawn from the ``right'' model should, on average, produce as
many fits with $\chi^2_\lambda$ above as below that for the fit to the
data: values near 50\% are achieved by the average dataset drawn
randomly from the model probability distribution.

\subsubsection{Model Results}

Examining Table 7, we find that the models with only three age bins
are significantly worse than those with four, and so are not discussed
further.  Model A, in which we assume solar metallicity throughout,
exclude the coolest stars and the most luminous one, and account for
the uncertainty in the extinction law, fits the data as well as it
does the average dataset drawn randomly from the model.  Model B,
which is identical to Model A except that we assume [Fe/H]$=-0.2$ for
ages $>$5 Gyr, is an unlikely fit to the data.  

As suggested by Figures~\ref{hrd0} and \ref{hrd2}, age and metallicity
are difficult to disentangle using only the tip of the AGB.  However,
our models do produce better fits to a SFH with purely solar [Fe/H]
from 0.01$-$12 Gyr. This may be understood through consideration of
Figure~\ref{color}. The {\it left} panel shows Model A compared with
the data, while the {\it right} panel shows Model B. The colors cyan,
magenta, yellow, and gray represent the model distributions scaled to
the best-fit SFR.  Darker regions indicate higher number density of
stars.  In the case of Model A, the footprints of the different age
bins align nicely with concentrations of observed data points.  In
Model B, the oldest age bin moves to higher temperature and
luminosity, becoming degenerate with the solar metallicity, younger
stars and forcing it to fit a larger number of points.  As a result,
Model B does not fit the coolest low luminosity stars well; indeed,
these stars are difficult to fit with anything other than a $>$ 12 Gyr
old solar metallicity model.  However, $>$ 12 Gyr--old stars with
[Fe/H]$\lesssim-0.6$ are less luminous than our \teff--dependent \mbol
\ limit. This means our SFH may not account for a potential very old
($>$ 12 Gyr old), metal--poor component which might have fewer
luminous AGB stars.

The resulting SFH for Models A\&B, showing the SFR in age bins of
10--100 Myr, 100 Myr -- 1 Gyr, 1--5 Gyr, and 5--12 Gyr, are given in
Figures~\ref{sfh} and \ref{sfhb}, respectively. The total initial mass represented by these SFHs are
shown in Figures~\ref{mass} and \ref{massb}, respecively. In Figure~\ref{sfh}, we have also plotted
a point corresponding to the model of \citet{krabbe95} which we take
as the average SFR over the last 10 Myr based on their burst model
which produces approximately 3200 M$_\odot$. 

\section{DISCUSSION}

A number of investigators have discussed the SFH in the
GC. \citet{lrt82} used the presence of young M supergiants to infer a
recent (\aple 10 Myr) burst of star formation. \citet{shks87} later
re--classified some of the same stars used by \citet{lrt82} as
luminous AGB stars, indicating intermediate ages (\apge 100 Myr) were
present as well. \citet*{ght94} discussed the SFH in terms of a
constant SFR and noted that such a model produces too few young blue
supergiants if it is adjusted to match the number of older late--type
stars. \citet{krabbe95} modeled the most recent epoch of star
formation in the nuclear cluster, producing the observed blue
supergiants with a burst of 10$^{3.5}$ M$_{\sun}$ over the last $\sim$ 7
Myr. This point is shown in Figure~\ref{sfh}. Apparently, the GC is
currently in a more quiescent state than in the recent
past. \citet{dav97} argued for an old population to dominate the
number counts within 6$''$ (0.23 pc) of the nuclear cluster based on near
infrared photometry. \citet{phil99} used surface brightness fitting
and photometry of individual stars to discuss the relative
contributions of young and old stars over a larger ($\sim
11'\times11'$) area. \citet{mez99} used the same data presented by
\citet{phil99} to further constrain the star formation
history. Recently, \citet{vl02} used photometric spectral energy
distributions of a large sample of stars in the Galactic bulge to
investigate the stellar populations there. Their results are
qualitatively similar to those presented here, though they investigate
a much larger area and have poor angular resolution in the central few
pc. BSD96 used near infrared spectra similar to those presented here
to identify young, intermediate age, and old stars. In the present
paper, we build on this earlier work by computing the SFH directly
from the observed stars using the \mbol \ and \teff \ determined from
the individual stellar spectra.

\subsection{Stellar Remnants}

Some fraction of the mass within the GC is due to massive stellar
remnants (neutron stars or black holes from initially massive stars)
which have migrated there from further out \citep{m93,meg00} by
dynamical friction with the low mass stellar population.  To the
extent that some of these massive objects were formed outside the 2 pc
radius covered by our data, that mass is unaccounted for in our SFH
model. \citet{m93} has computed migration times as a function of mass
and finds a remnant of 10 M$_{\sun}$ could migrate in to the center
from about 4 pc radius in a Hubble time. \citet{meg00} find a similar
result. The most massive remnants, black holes, will be able to
migrate inward from the largest radius. Their large mass and volume
over which they are drawn mean they will dominate this extra mass
component \citep{m93}.

\citet{m93} explored a range of models and found that up to $\sim$
10$\times$10$^{6}$~M$_{\sun}$ in massive remnants might have migrated
by dynamical friction to the central pc (Morris' models for similar
IMFs as used here produce masses \aple 5$\times$10$^{6}$~M$_{\sun}$).
This mass, which lies within a radius of 0.8 pc, overestimates the
dynamical mass including the central black hole for the most detailed
models mentioned above (by about a factor of two).  \citet{meg00}
estimate that 25,000 7 M$_{\sun}$ black hole remnants could have
settled in the central pc over a Hubble time; these remnants would
still be in the stellar cluster because the depletion timescale for
capture by the central black hole is 30 Gyr.  Given the somewhat lower
estimate of remnant mass due to migration by \citet{meg00} and the
estimate of $\sim$ 2 $\times$10$^{6}$~M$_{\sun}$ in a luminous stellar
cluster by \citet{scho02} within a 2 pc radius of the center, we expect
on order of at least a few times 10$^{6}$ M$_{\sun}$ in stars to have
formed in the inner 2 pc radius. If the Galaxy is 10 Gyr old, then this
implies a time averaged SFR of \apge \ 2$\times$10$^{-4}$
M$_{\sun}$~yr$^{-1}$.

\subsection{Depletion of CO Absorbers in the Central Parsec}

\citet{smbh90} showed that the CO absorption strength seen through
apertures centered on {\it unresolved} starlight was weaker inside a
radius of 15$''$ (0.58 pc) than outside this radius; see also \citet{genzel96}.
We now know the weakness of the CO feature is due in part to dilution
by very young stars \citep[ for example]{genzel96}. \citet{genzel96}
also showed that the brightest resolved sources with strong CO
(i.e. luminous AGB stars or supergiants) were absent from the inner
5$''$ (0.19 pc). Both \citet{smbh90} and \citet{genzel96} concluded it was
possible that the atmospheres of such stars might be destroyed by
collisions with lower mass stars in the GC leading to a deficiency of
CO--strong stars. \citet{bailey98} made detailed calculations of
collision probabilities in the GC and concluded that collisions
between giants and lower mass stars was unlikely to explain the
putative missing stars because such collisions (which form a common
envelope system) were ultimately ineffective in expelling the giant's
envelope on a time scale shorter than the evolutionary time
scale. However, \citet{davies98} did find that collisions between
giants and {\it binaries} might be effective in removing the giant
atmospheres in a short enough time to be observable (again through the
development of common--envelope systems). 
In any case, the deficiency
of resolved AGB or M supergiant stars is actually well concentrated to
the center \citep[$R$ \aple 5$'' = 0.2$ pc]{genzel96} and should thus not 
affect the SFH estimates for the larger area studied in this work 
($R$ $\sim$ 2$'$).

A consequence of the migration of massive remnants into the central
pc is the relaxation of the resulting dark cluster with the lower
mass stars which exist there. The model of \citet{meg00} predicts that
stars older than a few Gyr will be pushed to larger radii, forming a
distribution with a larger core radius (1--2 pc) and lower core
density than they would otherwise have. 
The implication of this prediction is that the present
data set would not be sensitive to the oldest epochs of star formation
in the GC if a significant fraction of the $\sim$ 1 M$_\odot$ 
tracers have been removed from the inner 2 pc, and so
might underestimate the total SFH. On the other hand, if the extent of
the dynamical redistribution of the low mass stars is toward the low end
of the range predicted by \citet[core radius $\sim$ 1 pc]{meg00} then
we would not expect this to be a large effect. The total amount of mass deduced
from dynamical models also constrains the SFH, and we discuss below the 
possibility that this constraint coupled with our models may provide evidence
that the dynamical friction effects are seen in the GC.

\subsection{The Star Formation History}

Figure~\ref{sfh} indicates significant on--going star formation in the
central few pc, but that the bulk of stars (roughly 75 $\%$ by mass)
formed at earlier times (Figure~\ref{mass}). This is in agreement with
earlier work based on near infrared number counts \citep{ght94,mez99},
and for the range of parameters discussed in \S3 and listed in
Table~7, this conclusion holds. For the oldest stars, we are sampling
just the very tip of the AGB, hence to observe any stars in such a
short lived phase requires a large mass to have originally formed. The
details change by roughly $\sim$ a factor of two depending on which
model SFH is chosen. The goodness of fit criterion, $P_{\lambda}$,
suggests the uniform metallicity case (Model A) is preferred. If true,
it could suggest that the nucleus formed largely from enriched
material produced in the early formation of the bulge. The purely
solar [Fe/H] is also consistent with the narrow distribution of [Fe/H]
from high resolution spectra \citep{ram00}. Though the range of ages
considered by \citet{ram00} is not as large as the data set presented
here, the high resolution data sample stars with ages up to $\sim$ 5
Gyr (Figure~\ref{hrd0}). In \S3, we noted that the current data set may
not be sensitive to metal poor populations older than 12 Gyr.

The total mass represented by the SFH in Figure~\ref{sfh} is shown in
Figure~\ref{mass} and is 9.9 $\pm$ 3.0 $\times$10$^6$ M$_{\sun}$ for
Model A. This is about three to six times larger than the most
detailed models (not counting the central black hole mass).

However, mass loss during the lifetime of stars from about one to 120
M$_{\sun}$ will reduce the cumulative final mass in the cluster. If
all stars with 120 M$_{\sun}$ $>$ M $>$ 1 M$_{\sun}$ are taken to have
their remnant mass at the present time, then, using the mapping of
initial mass to final mass given by \citet{m93}, the
present mass in stars is reduced by about 68 $\%$ (i.e. we infer the
present mass in stars remaining in the cluster to be 0.32 $\times$ the
total mass formed over all times) to a total of 3.2 $\pm$ 1
$\times$10$^6$ M$_{\sun}$ which is consistent with the dynamical
models cited in \S4.2.1. The total present day mass from the SFH depicted 
in Figure~\ref{sfh}, but including
mass loss, is shown in Figure~\ref{mass}. 
We have implicitly assumed all the tracers of
the SFH lie within a true radius of 2 pc, but they are actually
distributed in a projected radius of 2 pc. Given the steepness of the
stellar cluster radial density distribution (\aple 0.5 pc core
radius), the overestimate is likely to be small. The mass lost
through stellar winds could be expelled from the region and/or
recycled into new generations of stars. This is the maximal mass loss
since not all stars are yet old enough to have reached their final
mass, though most of the star formation has occurred at earlier times.

\citet{m93} argued that the IMF in the CMZ should be slanted toward
higher masses than the \citet{s55} mass function. \citet{fig99}
derived a flatter mass function in the nuclear young cluster, the
Arches, 30 pc from the GC. However, it is not clear if this is
representative of the initial mass function or dynamical effects
\citep{pz02}. An IMF slanted toward higher masses 
increases the total mass derived here since the IMF must
still produce the same number of low mass stars which form the
majority in Figure~\ref{hrd0}. For the same low mass cut--off of 0.7
M$_\odot$, the flatter IMF model yields a total mass of 16.4 $\pm$ 5.2
$\times$ 10$^{6}$ M$_\odot$. This model loses more mass through stellar winds 
resulting in a present day mass of 3.9 $\pm$ 1.2 $\times$ 10$^{6}$ M$_\odot$.
This is not too different than Model A discussed above. Higher values of the 
lower mass cut--off can reduce the total mass formed and thus the present day 
mass, but the cut--off is already near the limit of the mass of stars which 
we observe in the HRD (about 1 M$_\odot$). Apart from this, our models are 
not very sensitive to the details of the IMF. 

Our derived SFH may provide evidence that a significant number of low
mass stars have been removed from the central few pc as suggested by
\citet{meg00}. Our models, which trace the initial mass formed,
require that we cut--off the lower end of the mass function (below 0.7
M$_\odot$) in order to produce a final mass which is consistent with
the dynamical mass in central few pc. It is possible that this mass is
actually formed, but subsequently removed by dynamical friction as
massive remnants migrate toward the center \citep{m93,meg00}. A
better test of the dynamical relaxation effect predicted by \citet[as
they point out]{meg00} is to compute the radial distribution of the
many fainter low mass stars. Alternately, the cut--off at 0.7 M$_\odot$ 
may represent a bias to higher masses forming in the GC as argued by \citet{m93} and \citet{fig99}. 

\begin{figure}
\plotone{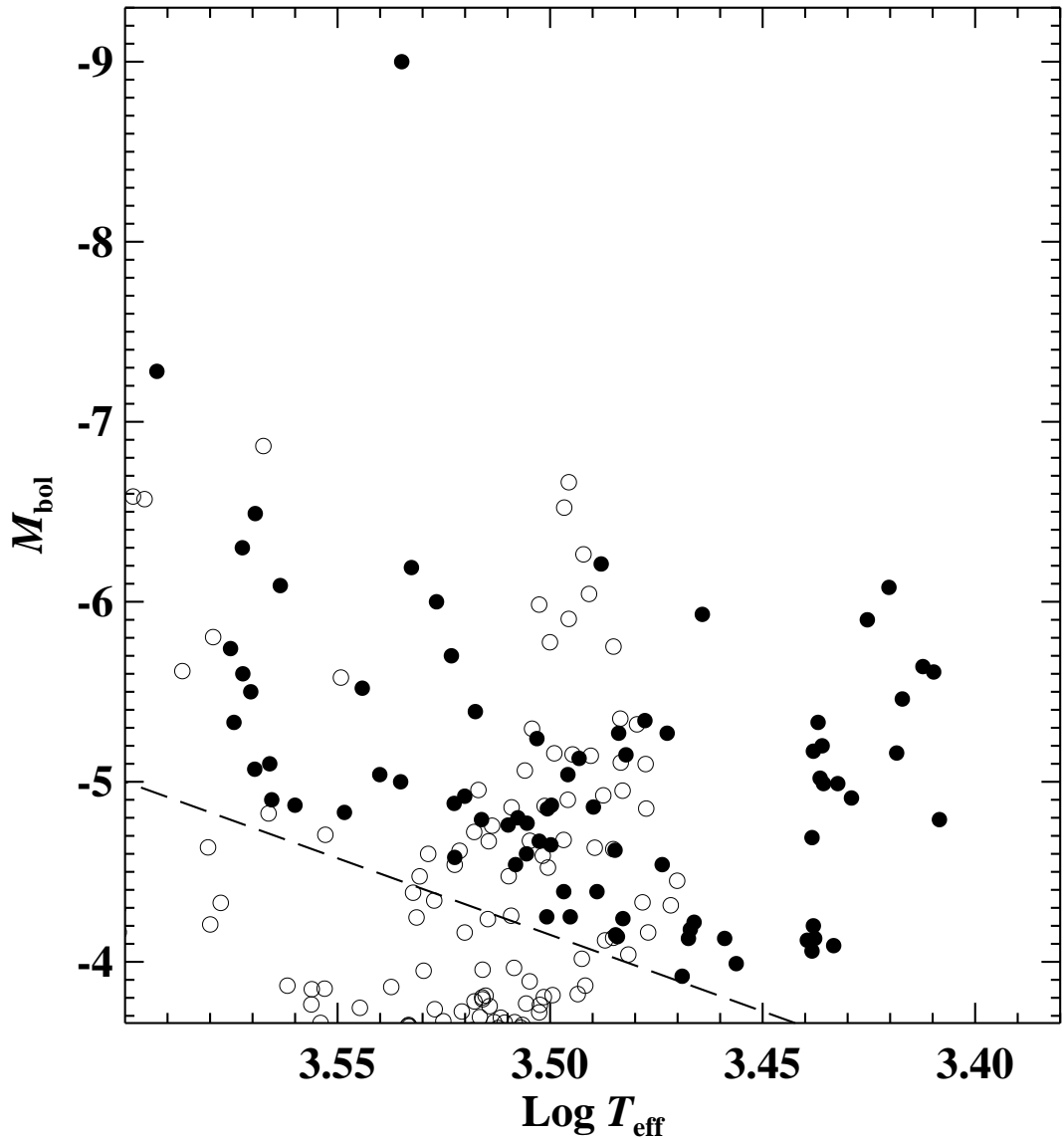} 
\caption{Hertzsprung--Russell diagram for the Galactic Center
(GC) stars ({\it filled circles}) compared to a model with constant
star formation rate ({\it open circles}). This figure demonstrates
that the models cover the same parameter space as the GC data,
including the pronounced intermediate age feature at \teff \ $\approx$
3200 K (log$_{10}$(\teff) $=$ 3.50), \mbol \ $\approx -$5.0, except
for the coolest GC stars (log$_{10}$(\teff) $>$ 3.45); see text.  The
model points include objects below our observed magnitude cut--off
(the region below the {\it dashed} line).
\label{testhrd}}
\end{figure}

\begin{figure}
\plotone{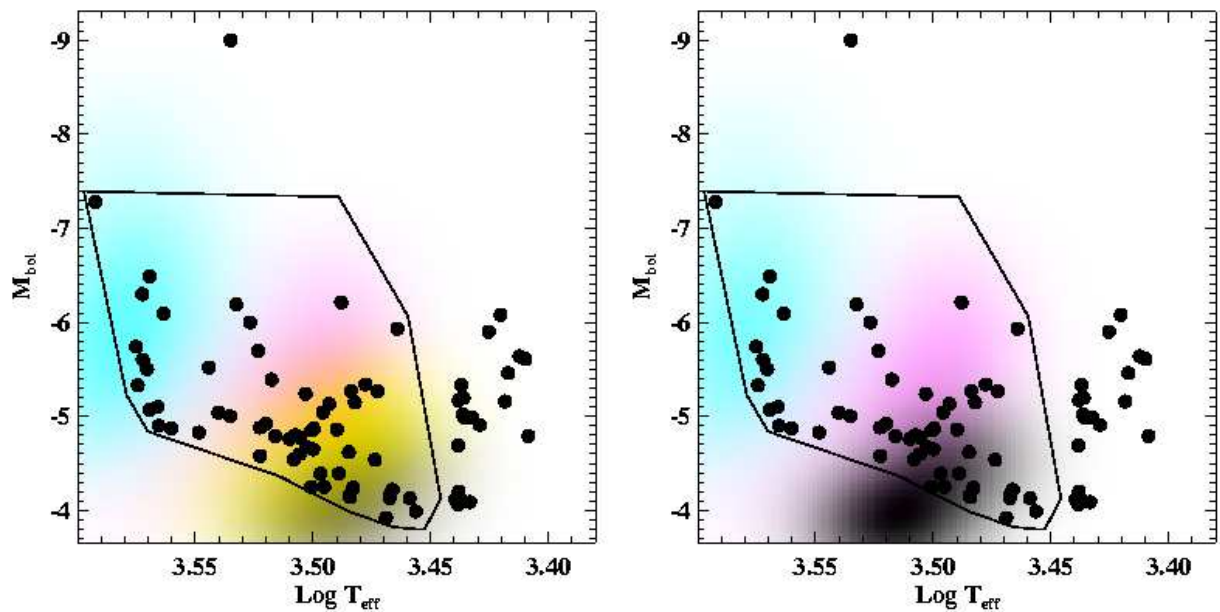}
\vspace{0.7 in} 
\caption{Comparison of the best fitting star formation history
with solar metallicity throughout ({\it left} panel, Model A, Table~7)
and with [Fe/H]=$-0.2$ for ages $\ge$ 5 Gyr ({\it right} panel, Model B,
Table~7). Darker regions represent higher number density. {\it Cyan}
corresponds to ages between 10 and 100 Myr, {\it
magenta} 100 Myr to 1 Gyr, {\it yellow} 1 Gyr to 5 Gyr, and {\it gray} to 5 Gyr to 12 Gyr. The purely
solar metallicity model fits the data better; see text. Model B ({\it
right} panel) has a best fit solution with no SF in the
3rd (1--5 Gyr) bin (this is why no {\it yellow} region appears;
see Figure~\ref{sfhb}).  However, a low metallicity ([Fe/H] \aple
-0.6), very old ($>$ 12 Gyr) component would not be detected by our
sample. The dark polygons in each panel represent the area of the
model and observational parameter space used in the fits. The coolest
stars are not accounted for by the models, and so were not used in the
fits (i.e. those stars outside the polygon), neither was the brightest
star, IRS7 (Age $<$ 10 Myr).
\label{color}}
\end{figure}

\begin{figure}
\plotone{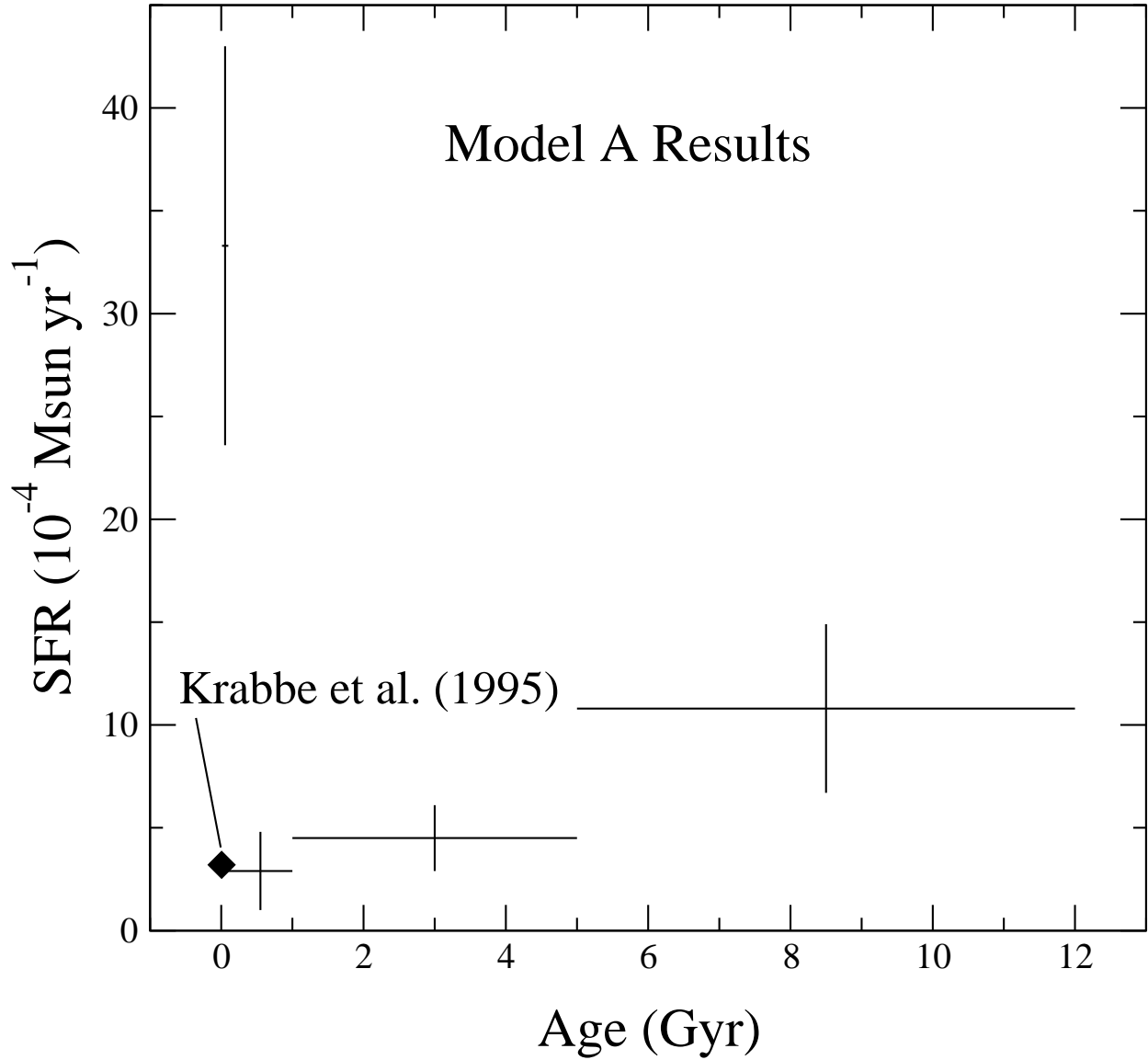}

\caption{Star formation history (SFH) for the Galactic center
(see Table~7). The crosses represent the results to the SFH fits to
the Hertzsprung--Russell Diagram (Figure~\ref{hrd0}) for Model A 
with solar [Fe/H] through out.
Model A provides the best fit;
see text. The age bins correspond to the horizontal width of the
crosses and are 10 Myr -- 100 Myr, 100 Myr -- 1 Gyr, 1 Gyr -- 5 Gyr,
and 5 Gyr -- 12 Gyr. The vertical height of each cross is the one
sigma error in the star formation rate for the respective bin. The
{\it filled} diamond represents the star burst model from
\citet{krabbe95} averaged over 10 Myr and is placed at 5 Myr along 
the age axis (i.e., it is the youngest point in the plot). 
\label{sfh}}
\end{figure}

\begin{figure}
\plotone{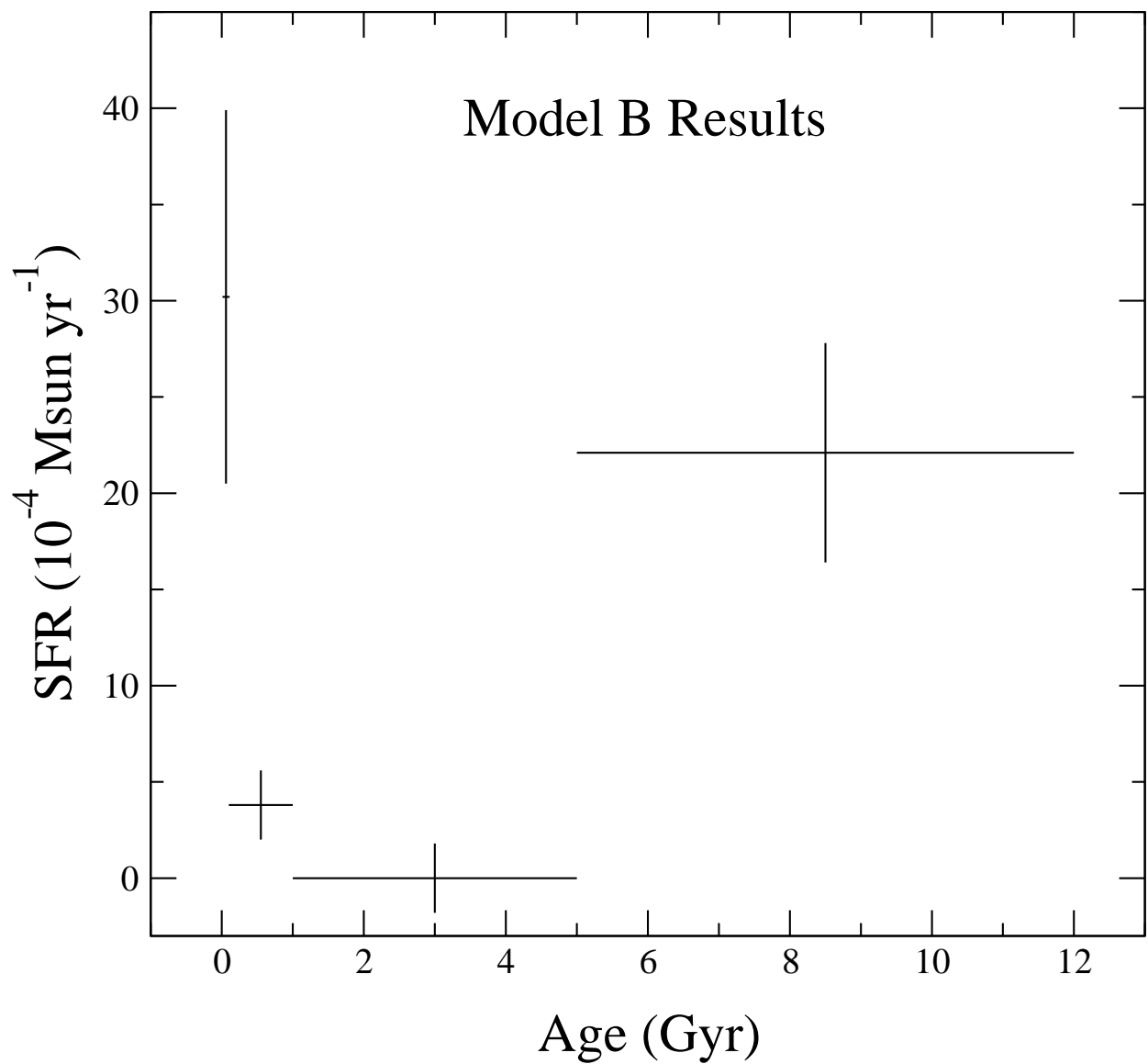}

\caption{
Same as Figure~\ref{sfh}, but for Model B 
with [Fe/H] $=-0.2$ in the oldest age bins. Model A (Figure~15$a$) 
provides a better fit; see text. 
\label{sfhb}}
\end{figure}

\begin{figure}
\plotone{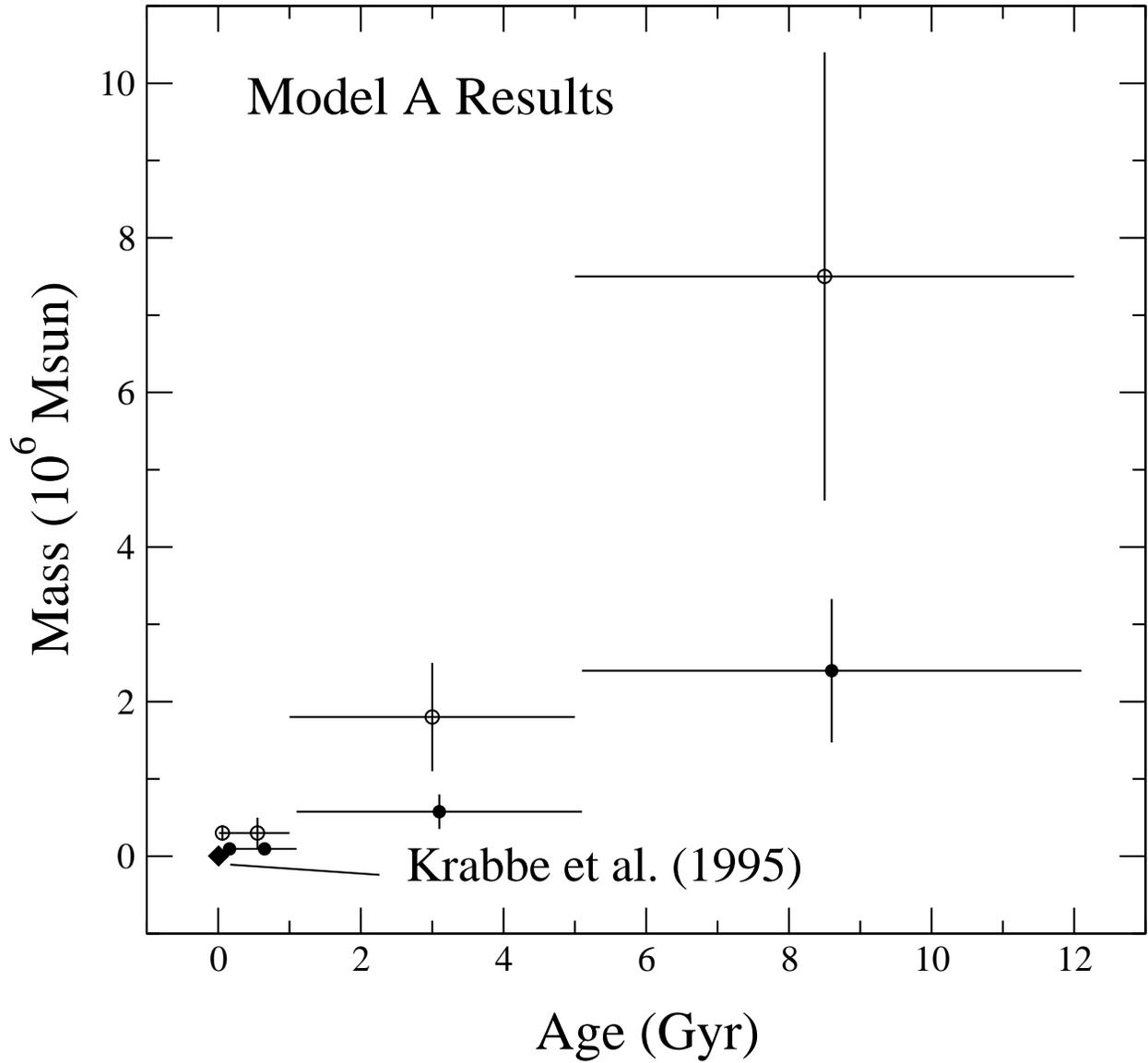}

\caption{Total mass ({\it open} circles) 
in stars formed in each corresponding age bin
(10 Myr -- 100 Myr, 100 Myr -- 1 Gyr, 1 Gyr -- 5 Gyr, and 5 Gyr -- 12
Gyr) for Model A of Figure~\ref{sfh}; see also Table~7. The
Krabbe et al. (1995) result for a $\sim$ 5 Myr old burst which
produces 3200 M$_\odot$ is shown.
The {\it filled} circles represent the present day mass for the same 
star formation history accounting for the mass
loss due to stellar winds. The total present day mass is consistent with the 
dynamical mass determinations; see text. The {\it filled} circles have been 
shifted by 0.1 Gyr in the Figure for clarity.
\label{mass}}
\end{figure}

\begin{figure}
\plotone{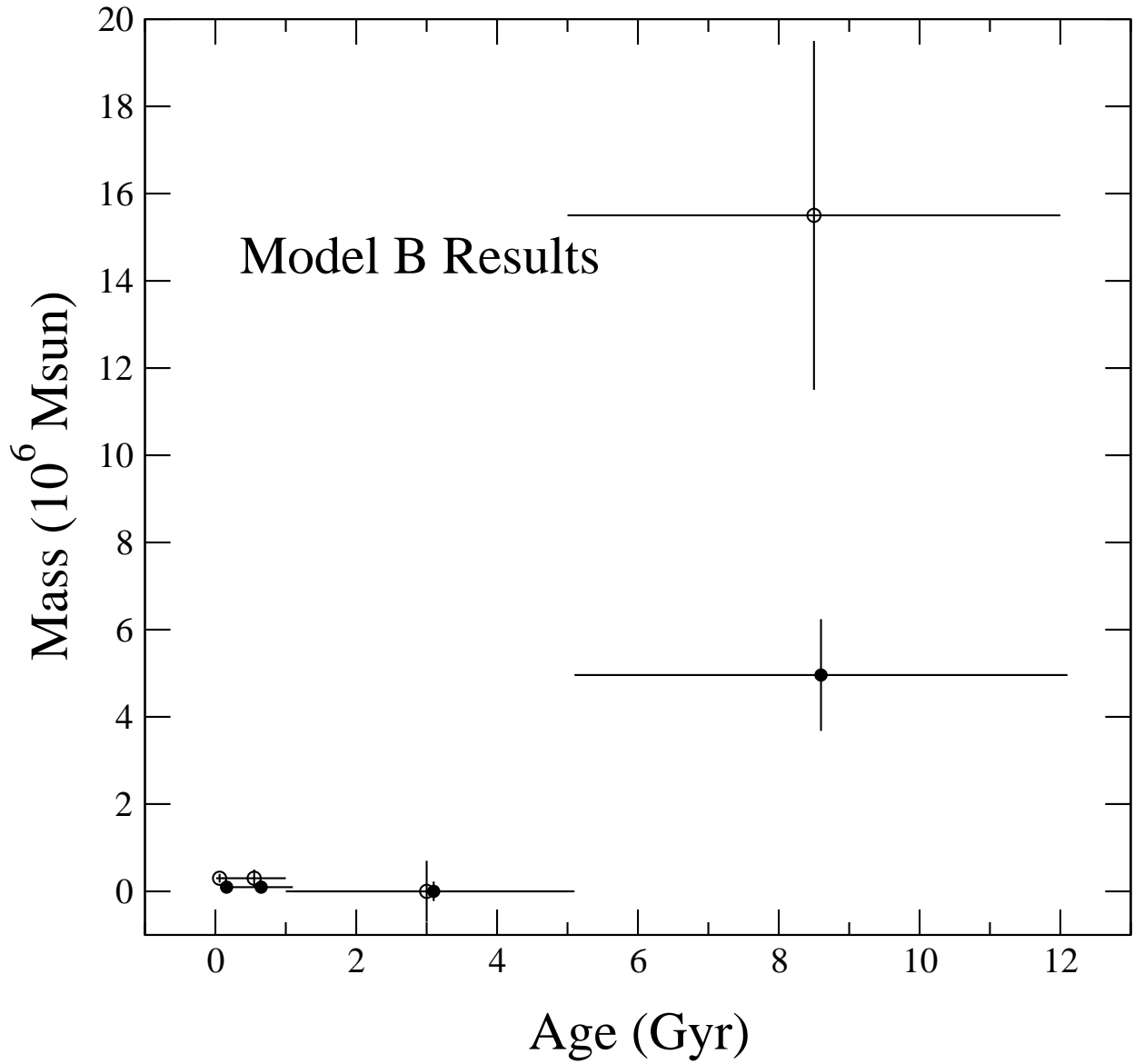}
\caption{Same as Figure~\ref{mass}, but for Model B results.
\label{massb}}
\end{figure}

Our technique of deriving \mbol \ and
\teff \ from low resolution spectra is preferred over
broad--band photometric analyses because of the large scatter in $A_K$
in the GC (BDS96) and the large variation in intrinsic colors of the
giant stars (which can be corrected for with the spectroscopically
derived \teff). Newly commissioned multi--object infrared
spectrometers on large telescopes are now available and will be used
in the near future to make these observations. Obtaining such
observations to 1.5--2 mag deeper than present will also improve the
SFH calculation (e.g., Figure~\ref{testhrd}) as the lower parts of the
older isochrones will be more densely populated. The observational
errors on \teff \ and \mbol \ still set a limit on the final age
resolution, however.

Our sample includes a number of luminous stars which require a
substantial amount of very recent star formation (10-100 Myr). The
high star formation rate at later times is reflected in
Figure~\ref{sfh} and Figure~\ref{testhrd}. The relatively large number
of luminous AGB and supergiant stars in this region of the HRD
requires significant recent star formation, perhaps in the form of
one or more concentrated bursts. Another way to see this is
by considering Figure~\ref{testhrd}. There is a relative paucity of
luminous stars in the part of the HRD covered by the youngest
isochrones for a constant star formation rate compared to the cooler,
less luminous AGB stars. Our models do not rule out high SFRs at early
times (in concentrated bursts); the lower mean SFRs result from the
wider age bins. The SFR in the youngest (and narrowest) age bin is
similar to the in fall rate of gas into the 2 pc molecular ring (see
\S1). This suggests star formation might be relatively efficient in the GC
($\sim$ 50--100$\%$), though the current SFR from \citet{krabbe95} is a factor
of 10 lower than in our most recent bin. 

In \S1, we noted that the CMZ was forming about 0.5 M$_\odot$
yr$^{-1}$. If we consider the SFR in the CMZ and GC per unit area
(taking a 200 pc radius disk--like distribution for the CMZ), then the
GC has formed stars at a prodigious rate over its history. The CMZ
normalized SFR is 4 $\times$ 10$^{-6}$ M$_\odot$ yr$^{-1}$
pc$^{-2}$. In the GC, the average SFR is $\sim$ 8 $\times$ 10$^{-4}$
M$_\odot$ yr$^{-1}$, and taking a radius of 2 pc, gives 6 $\times$
10$^{-5}$ M$_\odot$ yr$^{-1}$ pc$^{-2}$ with a peak of 
2.6 $\times$ 10$^{-4}$ M$_\odot$ yr$^{-1}$ pc$^{-2}$.

\citet{sjw99} have used luminous OH/IR (maser sources) stars as
tracers of star formation on larger scales (up to 50 pc) in the GC
region. They find evidence of significant star formation at an epoch
\apge \ 1 Gyr ago. \cite{ngd96} and \citet{whm98} also find evidence
for massive AGB, hence intermediate age, stars indicating significant
star formation on a similar time and spatial scale. \citet{ftk99}
studied the $K-$band number counts of stars in the inner Galaxy and
found that the younger (i.e. intermediate age) population may extend
out to a degree from the nucleus.  There is a feature in the HRD of
Figure~\ref{hrd0} which coincides with this age range. It is the
relatively dense group of stars centered at \teff \ $=$ 3300 K, \mbol
\ $=$ $-$5.0. This feature represents significant intermediate age star
formation ($\sim$ 1--2 Gyr), though not a significant fraction of the
total mass. The number of stars which trace out tracks near 1 Gyr in
Age is suggestive of a true ``burst,'' though our models do not have
the time resolution to conclusively limit the duration of this star
formation activity. \citet{vl02} have discussed the
properties of the stellar populations in the inner Galaxy using
broad--band photometric indices from the DENIS and ISOGAL
surveys. These surveys, whose angular resolution is more appropriate for
studies on large scales, give results which are broadly consistent
with those presented here. In particular, they find that the bulk of stars in
the inner Galaxy are old and not metal poor, that there has been
significant star formation at intermediate ages, and that current star
formation rates are relatively high. The correspondence of star
formation tracers in the central parsecs and on larger scales suggests
that star formation in the GC may be influenced by processes in the
inner Galaxy at large. The supply of gas to the central few pc may be
linked to the stellar bar which is thought to be a mechanism to funnel
star forming material to the inner Galaxy (see \S1).

\section{SUMMARY}

We have presented a Hertzsprung--Russell diagram (HRD) for a sample of
79 cool and luminous M type stars in the central few pc of the
Galaxy. The sample is based on a magnitude limited $K-$band data set
presented by \citet{bds96}. The \teff \ and \mbol \ were derived from
CO and \h2o \ molecular absorption features in $\lambda/\Delta\lambda
\sim$ 550 -- 1200 $H$ and $K-$band spectra. 

The HRD was used to derive the star formation history for the
Galactic center. Our sample of stars is too small to independently 
constrain all the
parameters in a detailed SFH (e.g, the slope of
the initial mass function, and the chemical enrichment history), thus
our SFH is not strictly unique. 
However, we find that the bulk of
stars in the Galactic center formed at early times (\apge 5 Gyr ago) for a
range of model parameters. There is also evidence for significant
recent star formation (\aple few Gyr ago). Such recent star formation
activity coincides in time with evidence from other evolved stars at
larger radii in the inner Galaxy ($>$ 50 pc), and suggests a
connection between star formation in the central pc and on larger
scales (presumably through gas input to the region). The age
resolution of our sample is not great due to observational errors on
the derived \mbol \ and \teff, the fact that the oldest isochrones are
not well separated along the asymptotic giant branch, and the
relatively small number of old, luminous stars, which trace the
majority of the derived mass.  Our best fitting models require a
cut--off in the IMF below a solar mass (at $\sim$ 0.7 M$_\odot$)
in order to produce a present--day mass in the central few pc which is
consistent with existing enclosed masses derived from dynamical
models. 
This ``cut--off'' might 
be evidence that mass segregation effects are at work in the GC, 
as has been predicted previously, or might instead point to a 
bias towards high mass star formation.
Finally, we find better 
fits to the data with models which have [Fe/H]=0.0 at all ages. This is 
consistent with earlier work at high spectral resolution which showed that 
stars between 10 Myr and $\sim$ 5 Gyr in the GC have solar metallicity.  

SVR and KS gratefully acknowledge support for this project from the NSF
through grants AST-9619230 and AST-0206331. This research has made use
of the SIMBAD database, operated at CDS, Strasbourg, France. We would
like to thank P. Martini, P. Romano, and J. An for helping obtain the
observations of Galactic center stars at CTIO and comparison stars at
MDM. Finally, we thank an anonymous referee whose comments and criticisms of our paper have led to its improvement.


\clearpage


\pagestyle{empty}
\begin{deluxetable}{lllrrlllll}
\rotate
\tablecaption{Comparison Star Table of Observations} 
\tablewidth{0pt}
\tabletypesize{\scriptsize}
\tablehead{ 
\colhead{HR} & 
\colhead{HD} &
\colhead{Other} &
\colhead{\teff \ (K)} &
\colhead{\mbol } &
\colhead{SpTyp} &
\colhead{\teff \ Reference\tablenotemark{a}} &
\colhead{\mbol \ Reference\tablenotemark{a}} &
\colhead{Obs Date} & 
\colhead{Instrument\tablenotemark{c}}
}
\startdata
\\
&       &       &       &     &  GIANTS          &       &       &       &  \\
\\
HR6418   & HD156283      & $\pi$ Her        & 4100  & $-$3.34         & K3II & Dy98  & Hip+La91      & 1999 May 05       & MOSAIC \\
HR4299   & HD095578      & 61 Leo        & 3700  & $-$3.04         & M0III     & Fe90  & Hip+La91      & 1999 March 03       & MOSAIC \\
 \nodata        & HD119667      & BD $-$02 3730   & 3700  & $-$5.10         & M1Swk         & SL90  & SL90  & 1999 March 01       & MOSAIC \\
HR5154   & HD119228      & IQ UMa        & 3600  & $-$3.60         & M2III    & Fe90  & Hip+La91      & 1999 March 02       & MOSAIC \\
\nodata         & HD147923      & BD +57 1671   & 3600  & $-$5.20         & M2S   & SL90  & SL90  & 1999 May 04       & MOSAIC \\
\enddata 

\tablecomments{The choice for inclusion with giants or supergiants for
stars with luminosity class II or with no luminosity class (the
S--stars) was made based on their measured CO strength; see \S3.2.1}

\tablenotetext{a}{References: Dy98 -- \citet{dy98}, Fe90 --
\citet{fe90}, SL90 -- \citet{sl90}, SL86 -- \citet{sl86}, Dy96 --
\citet{dy96}, Ram97 -- \citet{ram97}, SL85 -- \citet{sl85}, Lu82a --
\citet{lu82a}, LB80 -- \citet{lb80}, CSB00 -- \citet{csb00}, Ri98 --
\citet{ri98}, Ba98 -- \citet{ba98}, La86 -- \citet{la86}, vB97 --
\citet{vb97}, Hip -- \citet{hip}, La91 -- \citet{la91}, Be98 --
\citet{be98}, Dumm98 -- \citet{dumm98}, Lu82b -- \citet{lu82b}, Hu78
-- \citet{hu78}, EFH85 -- \citet*{efh85}, MJE95 -- \citet{mje95}, BL96
-- \citet{bl96}, vL97 -- \citet{vl97}}

\tablenotetext{b}{Analysis uses spectrum for
this star presented by \citet{bsd96} or \citet{lv92}.}

\tablenotetext{c}{IRS (CTIO 4m): $\lambda/\Delta\lambda = 560$,
OSIRIS (CTIO 4m): $\lambda/\Delta\lambda = 1200$, MOSAIC (MDM 1.3m): 
$\lambda/\Delta\lambda = 750$}

\tablenotetext{d}{The \mbol \ for R~Hya is more luminous than the
predicted maximum for AGB stars given by \citet{p70}; see also the discussion
by \citet{bsd96}. However the 1--$\sigma$ uncertainty on
its distance encompasses the luminosity limit.}

\end{deluxetable}

\pagestyle{empty}
\begin{deluxetable}{lrrrrrrrlll}
\rotate
\tablecaption{Galactic Center Stars} 
\tablewidth{0pt}
\tabletypesize{\scriptsize}
\tablehead{ 
\colhead{Name\tablenotemark{a}} & 
\colhead{Number\tablenotemark{b}} &
\colhead{Offset RA ($''$)\tablenotemark{c}} & 
\colhead{Offset Dec ($''$)\tablenotemark{c}} &
\colhead{$K$\tablenotemark{d}} &
\colhead{$J-K$\tablenotemark{d}} &
\colhead{$H-K$\tablenotemark{d}} &
\colhead{Obs Date / Ref} &
\colhead{Instrument\tablenotemark{e}} &
\colhead{Notes}
}
\startdata
        &1&$-$40.1  &$-$8.0   &10.40$\pm$0.08&\nodata&2.54$\pm$0.11& 2000 July 22        &OSIRIS & \\
        &2&$-$38.3  &$-$22.6  &10.27$\pm$0.05&\nodata&2.51$\pm$0.07& 2000 July 22        &OSIRIS & \\
        &4&$-$34.4  &$-$18.7  &10.36$\pm$0.04&6.85$\pm$0.07&2.74$\pm$0.06& 2000 July 22        &OSIRIS & \\
        &5&$-$29.4  &$-$23.2  &10.26$\pm$0.03&6.61$\pm$0.06&2.62$\pm$0.05& 2000 July 22        &OSIRIS & \\
        &6&$-$28.5  &$-$40.4  &9.67$\pm$0.03&\nodata&3.02$\pm$0.06& 1998 May 16        &IRS    & \\
        &7&$-$26.9  &18.9&9.80$\pm$0.03&7.15$\pm$0.06&2.76$\pm$0.05& 1998 May 15        &IRS    & \\
        &8&$-$24.0  &18.4&10.24$\pm$0.06&6.65$\pm$0.07&2.51$\pm$0.07& 1998 May 18        &IRS    & \\
        &9&$-$23.0  &16.4&9.92$\pm$0.04&6.29$\pm$0.05&2.42$\pm$0.05& 1998 May 17        &IRS    & \\
\enddata
\end{deluxetable}

\pagestyle{empty}
\begin{deluxetable}{lllrrrr}
\tablecaption{Comparison Star Indices} 
\tablewidth{0pt}
\tabletypesize{\scriptsize}
\tablehead{
\colhead{HR} & 
\colhead{HD} &
\colhead{Other} & 
\colhead{SpTyp} &
\colhead{\teff} &
\colhead{CO ($\%$)\tablenotemark{a}} &
\colhead{\h2o \ ($\%$)\tablenotemark{a}}
}
\startdata
&  & GIANTS & &&& \\
\\
HR6418	&	HD156283       	&	$\pi$-Her 	&	 K3II 	&	4100	&	13 $\pm$ 0.7     	&	1 $\pm$ 3 \\
HR4299  	&	HD095578       	&	61-Leo    	&	 M0III     	&	3700	&	15 $\pm$ 0.4    	&	2 $\pm$ 3 \\
...     	&	HD119667       	&	BD-023726 	&	 M1Swk         	&	3700	&	17 $\pm$ 0.5    	&	1 $\pm$ 3 \\
HR5154  	&	HD119228       	&	IQ-UMa    	&	 M2III    	&	3600	&	16 $\pm$ 0.5     	&	$-$1 $\pm$ 3 \\
...     	&	HD147923       	&	BD+571671 	&	 M2S   	&	3600	&	17 $\pm$ 0.5    	&	0 $\pm$ 3 \\
...     	&	...    	&	BD+062063 	&	 M3S   	&	3550	&	15 $\pm$0.5    	&	0 $\pm$ 3 \\
\enddata 

\tablenotetext{a}{Indices are $\%$ absorption defined as 100$\times(1
- Flux/Cont)$ where $Flux$ and $Cont$ are the integrated fluxes in
0.015 \mic \ bands in the spectra. For CO, the $Flux$ band is centered
at 2.302 \mic, and the $Cont$ band is taken at 2.284 \mic. For \h2o,
$Flux$ is for a band centered at 2.0675 \mic, and $Cont$ is the same
but derived from a fit to the spectral continuum; see text.  The
uncertainty in the measured \h2o \ strength is taken from the 
scatter in the measurements of IIIs and Is; see text.}

\tablenotetext{b}{Continuum may have been affected by data taking
procedure; see text. Strong expected \h2o \ absorption not evident.}

\tablenotetext{c}{$K-$band spectrum only, continuum fit made in the 
same way as for stars with $H$ and $K$, but only fit in the region
between 2.20 and 2.29 \mic.}

\end{deluxetable}

\pagestyle{empty}
\begin{deluxetable}{lccrrrr}
\tablecaption{Galactic Center Stars Younger than 10 Myr\tablenotemark{a}} 
\tablewidth{0pt}
\tabletypesize{\scriptsize}
\tablehead{ 
\colhead{Name\tablenotemark{b}} & 
\colhead{Number\tablenotemark{c}} &
\colhead{\teff / SpType\tablenotemark{d}} &
\colhead{$K_{\rm \circ}$\tablenotemark{e}} &
\colhead{$A_K$\tablenotemark{e}} &
\colhead{$M_K$\tablenotemark{f}} &
\colhead{$M_{\rm bol}$ \tablenotemark{d}}
}
\startdata
IRS 6E & 38& WC9 & 3.77 $\pm$ 0.28 &  6.28 $\pm$ 0.28 & $-$10.75& \\
IRS 34 & 46&WC9& 6.47 $\pm$ 0.12 &  4.01 $\pm$ 0.22& $-$8.05& \\
IRS 13E& 52&29000& 6.85 $\pm$ 0.18 &  2.97 $\pm$ 0.12 & $-$7.67& $-$11.1 \\
IRS 3  & 53&featureless& 6.78 $\pm$ 0.12 &  4.38 $\pm$ 0.11& $-$7.74 &\\
IRS 29N& 56&WC9& 3.56 $\pm$ 0.25 &  6.40 $\pm$ 0.10 & $-$10.96&\\
IRS 7&   66& 3430 $\pm$ 240  &  2.92 $\pm$ 0.16 & 3.48 $\pm$ 0.09&$-$11.60&$-$9.0 \\
IRS 16SW&71&24000& 7.00 $\pm$ 0.08 &  2.60 $\pm$ 0.06 & $-$7.52&$-$11.3 \\
MPE 1.6-6.8 &77&WC9& 6.65 $\pm$ 0.11 &  3.28 $\pm$ 0.08 & $-$7.87 &\\
\enddata

\tablenotetext{a}{Only stars with $K_\circ$ $\le$ 7.0 from the list
of \citet{bsd96} are included in this table.}

\tablenotetext{b}{Object names are as given by \citet{bn75},
\citet{l82}, \citet{tol89}, \citet{krabbe95} and \citet{bds96}}

\tablenotetext{c}{Number corresponds to sequence in
\citet{bds96}. Data tables in \citet{bds96} were truncated at \#~147.}

\tablenotetext{d}{\teff / \mbol \ for hot stars IRS 16NE, 16SW, and 13E from \citet{naj97}. \teff / \mbol \ for IRS 7 and VR5--7 from present work. Spectral types (SpType) from reference listed in Table 2.}

\tablenotetext{e}{$K_{\rm \circ}$ ($= K - A_K$) and $A_K$ taken from
\citet{bds96}. Stars \# 53, 78, and 81 use new unpublished
$H-$band magnitudes which were unavailable to \citet{bds96}.
$K_{\rm \circ}$ and $A_K$ for VR5--7 derived using photometry in Table 2.}

\tablenotetext{f}{$M_K$ is derived from $K_{\rm \circ}$ 
(see Blum et al. 1996$a$) and a distance modulus of 14.52 (8 kpc, Reid 1993).}

\end{deluxetable}

\pagestyle{empty}
\begin{deluxetable}{lccc}
\tablecaption{Completeness and Characteristics of the GC Sample} 
\tablewidth{0pt}
\tabletypesize{\scriptsize}
\tablehead{ 
\colhead{$K_{\circ}$\tablenotemark{a}} & 
\colhead{Percent Complete\tablenotemark{b}}&
\colhead{Number Observed}&
\colhead{Number in $K_{\rm \circ}$ LF}
}
\startdata
    $  K_{\circ}  <$ 4.0 & 100$\%$ &3  & 3\\
4.0 $< K_{\circ}  <$ 4.5 & 100$\%$ &0  & 0\\
4.5 $< K_{\circ}  <$ 5.0 & 100$\%$ &7  & 7\\
5.0 $< K_{\circ}  <$ 5.5 & 50$\%$  &5  & 10\\
5.5 $< K_{\circ}  <$ 6.0 & 85$\%$  &17 &20\\
6.0 $< K_{\circ}  <$ 6.5 & 78$\%$  &18 &23\\
6.5 $< K_{\circ}  <$ 7.0 & 55$\%$  &40 & 73\\
 & & &\\
Total (cool $+$ hot/young)&        &90 & 136\\ 
 & & &\\
Stars by type: & & &\\
Number of GC I & & & 15\tablenotemark{c}\\
Number of GC III & & & 43 \\
Number of GC LPV & & & 20 \\
 Stars Younger than 10 Myr & & &12 \tablenotemark{c}\\
 Stars not Observed       & & & 46\\
 & & &\\
Total                    &         &   & 136
\enddata

\tablenotetext{a}{The original sample is selected from the complete
list of stars with $K_{\circ}$ $<$ 7.0 mag in the central few pc of
the Galaxy presented by \citet{bds96}.}

\tablenotetext{b}{Percentage of stars observed spectroscopically
from the list of \citet{bds96} which includes known emission--line
stars or featureless young stars; see text and Tables~2 and 4.}

\tablenotetext{c}{IRS~7, M I, is counted among the young stars. VR5--7
of the Quintuplet cluster is not counted in this table, nor is \#72 which
is listed in Table 2 and 6.}

\end{deluxetable}

\pagestyle{empty}
\begin{deluxetable}{lrrrcccccccr}
\rotate
\tablecaption{Galactic Center Star Properties} 
\tablewidth{0pt}
\tabletypesize{\scriptsize}
\tablehead{ 
\colhead{Name} & 
\colhead{Number} &
\colhead{CO ($\%$)} &
\colhead{\h2o \ ($\%$)} & 
\colhead{Luminosity Class\tablenotemark{b}} &
\colhead{\teff\tablenotemark{c}} &
\colhead{$K_{\rm \circ}$\tablenotemark{d}} &
\colhead{$M_K$\tablenotemark{e}} &
\colhead{$A_K$\tablenotemark{d}} &
\colhead{BC$_K$\tablenotemark{f}} &
\colhead{$\Delta A_K$\tablenotemark{g}} &
\colhead{\mbol\tablenotemark{h}}
}
\startdata
   	&	1	&	21   $\pm$0.6   	&	$-$2	&	I   	&	3710	$\pm$	130	&	6.85	$\pm$	0.12	&	$-$7.67	&	3.55  $\pm$ 0.09   	&	2.6	&	0.0	&	$-$5.07	$\pm$	0.42	\\
   	&	2	&	20   $\pm$0.8   	&	17	&	LPV?   	&	2740	$\pm$	200	&	6.76	$\pm$	0.12	&	$-$7.76	&	3.51  $\pm$ 0.11   	&	3.2	&	0.5	&	$-$4.06	$\pm$	0.42	\\
   	&	4	&	26   $\pm$0.9   	&	16	&	LPV?   	&	2750	$\pm$	200	&	6.70	$\pm$	0.10	&	$-$7.82	&	3.67  $\pm$ 0.06   	&	3.2	&	0.5	&	$-$4.12	$\pm$	0.41	\\
   	&	5	&	24   $\pm$0.9   	&	10	&	III   	&	2930	$\pm$	150	&	6.74	$\pm$	0.06	&	$-$7.78	&	3.51  $\pm$ 0.05   	&	3.2	&	0.4	&	$-$4.18	$\pm$	0.40	\\
   	&	6	&	17   $\pm$2.1   	&	32	&	LPV?   	&	2610	$\pm$	200	&	5.36	$\pm$	0.11	&	$-$9.16	&	4.31  $\pm$ 0.11   	&	3.2	&	0.5	&	$-$5.46	$\pm$	0.41	\\
   	&	7	&	23   $\pm$1.9   	&	15	&	III   	&	3040	$\pm$	250	&	5.97	$\pm$	0.06	&	$-$8.55	&	3.83  $\pm$ 0.05   	&	3.1	&	0.3	&	$-$5.15	$\pm$	0.40	\\
   	&	8	&	22   $\pm$1.2   	&	21	&	LPV?   	&	2710	$\pm$	200	&	6.73	$\pm$	0.08	&	$-$7.79	&	3.50  $\pm$ 0.06   	&	3.2	&	0.5	&	$-$4.09	$\pm$	0.41	\\
   	&	9	&	20   $\pm$0.9   	&	10	&	III   	&	3280	$\pm$	150	&	6.63	$\pm$	0.06	&	$-$7.89	&	3.29  $\pm$ 0.05   	&	2.9	&	0.2	&	$-$4.79	$\pm$	0.40	\\
\enddata

\tablenotetext{a}{Indices are $\%$ absorption defined as 100$\times(1
- Flux/Cont)$ where $Flux$ and $Cont$ are the integrated fluxes in
0.015 \mic \ bands in the spectra. For CO, the $Flux$ band is centered
at 2.302 \mic, and the $Cont$ band is taken at 2.284 \mic. For \h2o,
$Flux$ is for a band centered at 2.0675 \mic, and $Cont$ is the same
but derived from a fit to the spectral continuum; see text. The uncertainty
in the measured \h2o \ strength (3$\%$) 
is taken from the scatter in the measurements
for comparison star IIIs and Is; see text.}

\tablenotetext{b}{The luminosity class is set by the CO index
and the \h2o \ index; see text.}

\tablenotetext{c}{\teff \ is derived from the CO index using
the relationship of CO vs \teff \ derived for the comparison
stars. See the discussion in the text and Figure~\ref{covteff}.}

\tablenotetext{d}{$K_{\rm \circ}$ ($= K - A_K$) and $A_K$ taken 
from \citet{bds96}.
For \#'s 1, 2, 6, 11, 40, 148, 162, and 170 $A_K$ were
derived in the same way, but using new unpublished $H$ magnitudes
as given in Table~2.}

\tablenotetext{e}{$M_K$ is derived from $K_{\rm \circ}$ and a 
distance modulus of 14.52 (8 kpc, Reid 1993).}.

\tablenotetext{f}{Bolometric correction to the $K_{\circ}$ magnitudes. The 
correction is a linear function of \teff \ for giants, and a constant for
LPV's and supergiants; see text and \citet{bsd96}.}

\tablenotetext{g}{Correction to $A_K$ to account for the range of intrinsic
color in the GC stars. $\Delta A_K = A_K$\citep{bds96} $- A_K$(Corrected). See text.}

\tablenotetext{h}{\mbol \ $=$ $K_{\circ} \ +$ BC$_K$ $ + \ \Delta A_K \
-$ DM, where DM is the distance modulus to the GC (14.52) for 8 kpc
\citep{r93}.}

\tablenotetext{i} {Star 72 is included because it was observed and analyzed at 
high spectral resolution by \citet{ram00}.
This star was not used in the star formation history calculation because 
its $K_{\rm \circ}$ is below the cut--off adopted; see \S3.3.}

\end{deluxetable}

\begin{deluxetable}{lcccccrrr}
\tabletypesize{\scriptsize}
\tablewidth{0pt}
\tablecaption{Summary of Models Fitted\label{table}}
\tablehead{
\colhead{Model} &
\colhead{$\chi_\lambda^2$} &
\colhead{$P_\lambda$} & 
\colhead{$M_T$} &
\colhead{$\sigma_{M_T}$} &
\colhead{Age} & 
\colhead{[Fe/H]} & 
\colhead{SFR} &
\colhead{ $\sigma_{\rm SFR}$} \\
\colhead{} &
\colhead{} &
\colhead{(\%)} & 
\colhead{(10$^6 M_\sun$)} &
\colhead{} &
\colhead{(Gyr)} & 
\colhead{} & 
\colhead{(10$^{-4} M_\sun$ yr$^{-1}$)} &
\colhead{} \\
}
\startdata
    Model A &  231.9 &  58.9 &   9.9 &    3.0 &    0.01 $-$ 0.10 &   0.0 &  33.3 &  9.7 \\ 
    &  &  &  &  &                                  0.10 $-$ 1.00 &   0.0 &   2.9 &   1.9 \\ 
 &  &  &  &  &                                     1.00 $-$ 5.00 &   0.0 &   4.5 &   1.6 \\ 
 &  &  &  &  &                                     5.00 $-$ 12.00 &   0.0 &  10.8 &   4.1 \\ 
    Model A, &  232.4 &  54.8 &   16.4 &   4.3 &   0.01 $-$ 0.10 &   0.0 &  28.3 &   8.5 \\ 
 ~~IMF slope varied&  &  &  &  &                   0.10 $-$ 1.00 &   0.0 &   3.4 &   2.4 \\ 
 &  &  &  &  &                                     1.00 $-$ 5.00 &   0.0 &   7.3  &   2.4 \\ 
 &  &  &  &  &                                     5.00 $-$ 12.00 &   0.0 &  18.4 &   7.3 \\
\enddata
\end{deluxetable}

\clearpage

\end{document}